\documentclass[aps,prb,twocolumn,superscriptaddress]{revtex4-2}
\usepackage{graphicx}
\usepackage{color,amssymb,amsmath}
\usepackage[normalem]{ulem}
\usepackage{hyperref}
\usepackage{bm}

\newcommand{\Vt}{V_{\mathrm{T}}}
\newcommand{\Vtl}{V_{\mathrm{TL}}}
\newcommand{\Vtr}{V_{\mathrm{TR}}}
\newcommand{\Vj}{V_{\mathrm{Switch}}}
\newcommand{\Vg}{V_{\mathrm{G}}}
\newcommand{\Vp}{V_{\mathrm{Probe}}}
\newcommand{\Il}{I_{\mathrm{L}}}
\newcommand{\Ir}{I_{\mathrm{R}}}
\newcommand{\PhiL}{\mathit{\Phi}_{\mathrm{L}}}
\newcommand{\PhiR}{\mathit{\Phi}_{\mathrm{R}}}
\newcommand{\Vsd}{V_{\mathrm{SD}}}

\newcommand{\phiL}{\phi_{\mathrm{L}}}
\newcommand{\phiM}{\phi_{\mathrm{M}}}
\newcommand{\phiR}{\phi_{\mathrm{R}}}

\newcommand{\El}{E_\mathrm{L}}
\newcommand{\Er}{E_\mathrm{R}}
\newcommand{\Deltait}{\mathit{\Delta}}
\newcommand{\Gammait}{\mathit{\Gamma}}
\newcommand{\vphiL}{\varphi_{\mathrm{L}}}
\newcommand{\vphiR}{\varphi_{\mathrm{R}}}

\begin{document}
\title{Phase-engineering the Andreev band structure of a three-terminal Josephson junction}

\author{M.\ Coraiola}
\affiliation{IBM Research Europe---Zurich, 8803 R\"uschlikon, Switzerland}

\author{D.\ Z.\ Haxell}
\affiliation{IBM Research Europe---Zurich, 8803 R\"uschlikon, Switzerland}

\author{D.\ Sabonis}
\affiliation{IBM Research Europe---Zurich, 8803 R\"uschlikon, Switzerland}

\author{H.\ Weisbrich}
\affiliation{Fachbereich Physik, Universit\"at Konstanz, D-78457 Konstanz, Germany}

\author{A.\ E.\ Svetogorov}
\affiliation{Fachbereich Physik, Universit\"at Konstanz, D-78457 Konstanz, Germany}

\author{M.\ Hinderling}
\affiliation{IBM Research Europe---Zurich, 8803 R\"uschlikon, Switzerland}

\author{S.\ C.\ ten Kate}
\affiliation{IBM Research Europe---Zurich, 8803 R\"uschlikon, Switzerland}

\author{E.\ Cheah}
\affiliation{Laboratory for Solid State Physics, ETH Z\"urich, 8093 Z\"urich, Switzerland}

\author{F.\ Krizek}
\altaffiliation[Present address: ]{Institute of Physics, Czech Academy of Sciences, 162 00 Prague, Czech Republic}
\affiliation{IBM Research Europe---Zurich, 8803 R\"uschlikon, Switzerland}
\affiliation{Laboratory for Solid State Physics, ETH Z\"urich, 8093 Z\"urich, Switzerland}

\author{R.\ Schott}
\affiliation{Laboratory for Solid State Physics, ETH Z\"urich, 8093 Z\"urich, Switzerland}

\author{W.\ Wegscheider}
\affiliation{Laboratory for Solid State Physics, ETH Z\"urich, 8093 Z\"urich, Switzerland}

\author{J.\ C.\ Cuevas}
\affiliation{Departamento de Física Te\'orica de la Materia Condensada and Condensed Matter Physics Center (IFIMAC), Universidad Aut\'onoma de Madrid, E-28049 Madrid, Spain}

\author{W.\ Belzig}
\affiliation{Fachbereich Physik, Universit\"at Konstanz, D-78457 Konstanz, Germany}

\author{F.\ Nichele}
\email{fni@zurich.ibm.com}
\affiliation{IBM Research Europe---Zurich, 8803 R\"uschlikon, Switzerland}

\date{October 25, 2023}

\begin{abstract}
In hybrid Josephson junctions with three or more superconducting terminals coupled to a semiconducting region, Andreev bound states may form unconventional energy band structures, or Andreev matter, which are engineered by controlling superconducting phase differences. Here we report tunnelling spectroscopy measurements of three-terminal Josephson junctions realised in an InAs/Al heterostructure. The three terminals are connected to form two loops, enabling independent control over two phase differences and access to a synthetic Andreev band structure in the two-dimensional phase space. Our results demonstrate a phase-controlled Andreev molecule, originating from two discrete Andreev levels that spatially overlap and hybridise. Signatures of hybridisation are observed in the form of avoided crossings in the spectrum and band structure anisotropies in the phase space, all explained by a numerical model. Future extensions of this work could focus on addressing spin-resolved energy levels, ground state fermion parity transitions and Weyl bands in multiterminal geometries.
\end{abstract}

\maketitle

\section*{Introduction}
In a normal conductor interfacing two or more superconductors, charge carriers at energies within the superconducting gap are confined by Andreev reflection processes occurring at the interfaces \cite{Andreev1964}. As a result, resonant sub-gap electronic excitations known as Andreev bound states (ABSs) arise in the normal region, enabling transport of a Josephson supercurrent between the superconducting terminals \cite{Beenakker1991, Furusaki1991}. These discrete ABS levels were proposed as a basis for quantum computing applications \cite{Desposito2001, Zazunov2003, Chtchelkatchev2003, Padurariu2010}. More recently, ABSs have been the subject of intense experimental investigation in various material platforms \cite{Pillet2010, Chang2013, Bretheau2013a, Bretheau2013b, Bretheau2017, VanWoerkom2017, Tosi2019, Nichele2020}, culminating in the coherent control of Andreev pair \cite{Janvier2015, Hays2018} and spin \cite{Hays2021, PitaVidal2023} qubits.
While these studies focused on Josephson junctions (JJs) with two superconducting terminals, where the ABS energies depend on a single superconducting phase difference, multiterminal JJs (MTJJs) have also emerged as a promising alternative. In the presence of $N \geq 3$ terminals, the ABS band structure spanned by the $N-1$ independent phase differences was predicted to exhibit a plethora of phenomena, including lifting of the spin degeneracy \cite{vanHeck2014}, ground state fermion parity transitions \cite{vanHeck2014}, Weyl singularities \cite{Yokoyama2015, Riwar2016, Eriksson2017, Meyer2017, Xie2017} and other topological properties \cite{Deb2018, Xie2019, Repin2019, Gavensky2019, Houzet2019, Klees2020}. Moreover, MTJJs were proposed to realise Andreev molecules \cite{Pillet2019, Kornich2019, Kornich2020, Pillet2020}---a system where single ABSs overlap and form hybridised energy levels---and explored as a platform to generate Cooper quartets \cite{Freyn2011, Jonckheere2013, Pfeffer2014, Cohen2018, Huang2022}. Extensive experimental work was also conducted on MTJJs in the presence of DC current bias \cite{Draelos2019, Graziano2020, Pankratova2020, Matsuo2022} and microwave irradiation \cite{Arnault2021}. While tunnelling spectroscopy of metallic three-terminal JJs (3TJJs) was performed previously \cite{Strambini2016}, establishing phase control over the superconducting proximity gap, further investigation is needed to understand the properties of ABS bands in multiterminal devices. The perspective of engineering synthetic band structures by exploiting the higher dimensionality of the phase space is particularly attractive, as it could enable effects unattainable in two-terminal geometries.

Here, we report on an experimental realisation of superconductor--semiconductor 3TJJs and study ABSs in the system with tunnelling spectroscopy. Owing to the independent control over two superconducting phase differences, we probe the Andreev band structure in the two-dimensional (2D) phase space and find signatures of hybridisation between highly transmissive ABSs, resulting in the formation of an Andreev molecule. Our measurements are supported by a theoretical model and demonstrate the feasibility of Andreev matter and phase-engineering of Andreev bands in hybrid nanostructures.

\begin{figure*}
	\includegraphics[width=\textwidth]{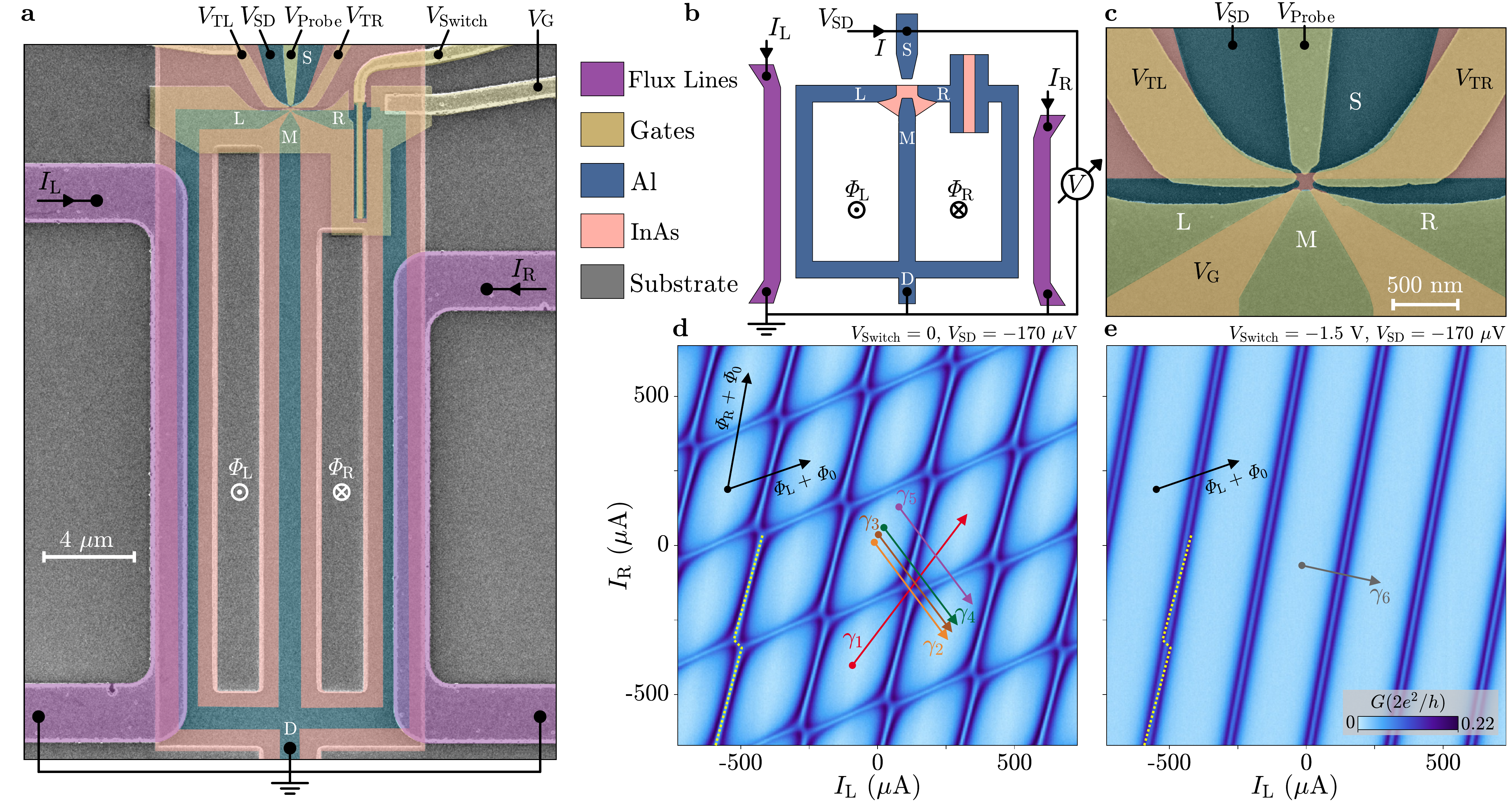}
	\caption{\textbf{Experimental setup and tunnelling conductance in the two-dimensional phase space.} 
		\textbf{a}, False-coloured scanning electron micrograph of a device identical to that under study, defined by selective removal of the Al (blue), exposing the semiconductor below (pink). Gates (yellow) and flux-bias lines (purple) were deposited on a uniform dielectric layer (not visible). Bias voltage $\Vsd$, gate voltages $V_\alpha$ ($\alpha \in \{ \mathrm{TL,TR,Probe,Switch,G} \}$), left (right) flux-bias line current $I_\mathrm{L(R)}$ and external magnetic flux threading the left (right) superconducting loop $\mathit{\Phi}_\mathrm{L(R)}$ are indicated.
		\textbf{b}, Schematic representation of the device and the measurement setup.
		\textbf{c}, Zoom-in of \textbf{a} near the three-terminal Josephson junction (3TJJ) region.
		\textbf{d},\textbf{e}, Differential conductance $G$ between tunnelling probe and 3TJJ measured as a function of the currents $\Il$ and $\Ir$ injected into the flux-bias lines, at fixed voltage bias $\Vsd = -170 ~\mu \mathrm{V}$. In \textbf{d} (\textbf{e}), the switch junction is in the ON (OFF) state, $\Vj = 0$ ($\Vj = -1.5~\mathrm{V}$). The directions of the black arrows indicate the periodicity axes, along which the external magnetic fluxes $\PhiL$ and $\PhiR$ vary independently. Each black arrow represents the addition of one superconducting flux quantum $\mathit{\Phi}_0 = h/2e$ (where $h$ is the Planck constant and $e$ the elementary charge) to the corresponding flux. The dotted yellow line follows a $\PhiL$-dependent resonance in \textbf{d} and is replicated in \textbf{e} to highlight the slope difference. The coloured arrows labelled $\gamma_\mathrm{1-6}$ define the linecuts shown in Fig.~\ref{fig2}.
	}
	\label{fig1}
\end{figure*}

\section*{Results}
\subsection*{Implementation of phase-controllable 3TJJ}
The device under investigation, shown in Fig.~\ref{fig1}a--c, was realised in an InAs/Al heterostructure \cite{Shabani2016, Cheah2023}. Selective etching of the Al layer defined three superconducting terminals (labelled L, M and R) coupled to a normal region, constituting a hybrid 3TJJ. The three terminals were connected to a common node (D) forming two closed loops; while leads L and M were directly connected to D via Al strips, a superconductor--normal--superconductor (SNS) JJ was integrated on terminal R. The junction, with a length of 40 nm and a width of 5 $\mu$m, was designed to have a critical current much larger than that existing between any pairs of L, M and R, hence the superconducting phase difference across the junction is neglected for the following discussion. A fourth superconducting lead (S) was employed as a probe to enable DC tunnelling spectroscopy of the sub-gap states in the 3TJJ. Metallic gate electrodes and flux-bias lines were patterned on top of an insulating layer uniformly deposited across the entire sample. The transmission between the probe and the 3TJJ was tuned via two gates energised by the common voltage $\Vt \equiv \Vtl = \Vtr$, which was set to $-1.07~\mathrm{V}$ to enter the tunnelling regime: in this configuration, the probe was weakly coupled to the 3TJJ and its influence on the rest of the circuit was limited. Gate tuning of the SNS junction enabled its operation as a switch, with the ON (OFF) state defined by $\Vj = 0$ ($\Vj = -1.5~\mathrm{V}$). This allowed for the connection or disconnection of terminal R from D, hence electrostatically selecting between a three-terminal (switch ON) and a two-terminal configuration (switch OFF). Two additional gate voltages were kept to $\Vp = 150~\mathrm{mV}$ and $\Vg = 50~\mathrm{mV}$. A current $I_\mathrm{L(R)}$ injected into the left (right) flux-bias line generated an external magnetic flux $\mathit{\Phi}_\mathrm{L(R)}$ threading the left (right) superconducting loop, thus tuning the phase difference between L (R) and M, and enabling control over the whole 2D phase space. As schematically depicted in Fig.~\ref{fig1}b, a DC voltage bias $\Vsd$ was applied to the probe S and the differential tunnelling conductance $G$ was measured between S and D with standard lock-in techniques. Experiments were performed in a dilution refrigerator with base temperature below $10~\mathrm{mK}$. Further details about materials, fabrication and measurement are provided in the Methods section.

\subsection*{Probing the Andreev band structure in the 2D phase space}
In Fig.~\ref{fig1}d,e, the voltage bias was set to $-170~\mu \mathrm{V}$ and the tunnelling conductance was measured as a function of the currents $\Il$ and $\Ir$ injected into the flux-bias lines, resulting in a scan over an extended region of the 2D phase space at constant energy. Resonances in conductance correspond to peaks in the density of states (DOS) of the normal region under study \cite{Pillet2010, Nichele2020} and represent ABSs in the 3TJJ. Each state is intersected twice per period at $\Vsd = -170~\mu \mathrm{V}$ (i.e., at an energy below its maximum), giving rise to the characteristic appearance of pairs of lines in these maps.
When the switch is set to the ON state (Fig.~\ref{fig1}d), we observe periodic features as a function of both $\Il$ and $\Ir$, attributed to the presence of two distinct ABSs whose energies disperse with $\PhiL$ and $\PhiR$, respectively. The cross-dependence between the flux-bias lines accounts for the finite slope of the $(\PhiL, \PhiR)$ axes, as indicated by the black arrows.
Remarkably, resonances associated to different ABSs are connected to each other in proximity of the intersections, forming closed diamond-like loops and avoided crossings at the corners of the diamonds. Each state undergoes a phase shift when intersecting the other, defining a zig-zag trend. Next, we set the switch junction voltage to $-1.5~\mathrm{V}$ (OFF, Fig.~\ref{fig1}e) and observe that the ABS resonances depending on $\PhiR$ disappear, while the complex 2D periodic pattern is transformed into a simple structure depending on a single flux. Notably, the slope of the $\PhiL$-dependent resonances differs between panels d and e (see dotted yellow lines), which is compatible with the periodic phase shift present only when the switch is ON.

\begin{figure}
	\includegraphics[width=\columnwidth]{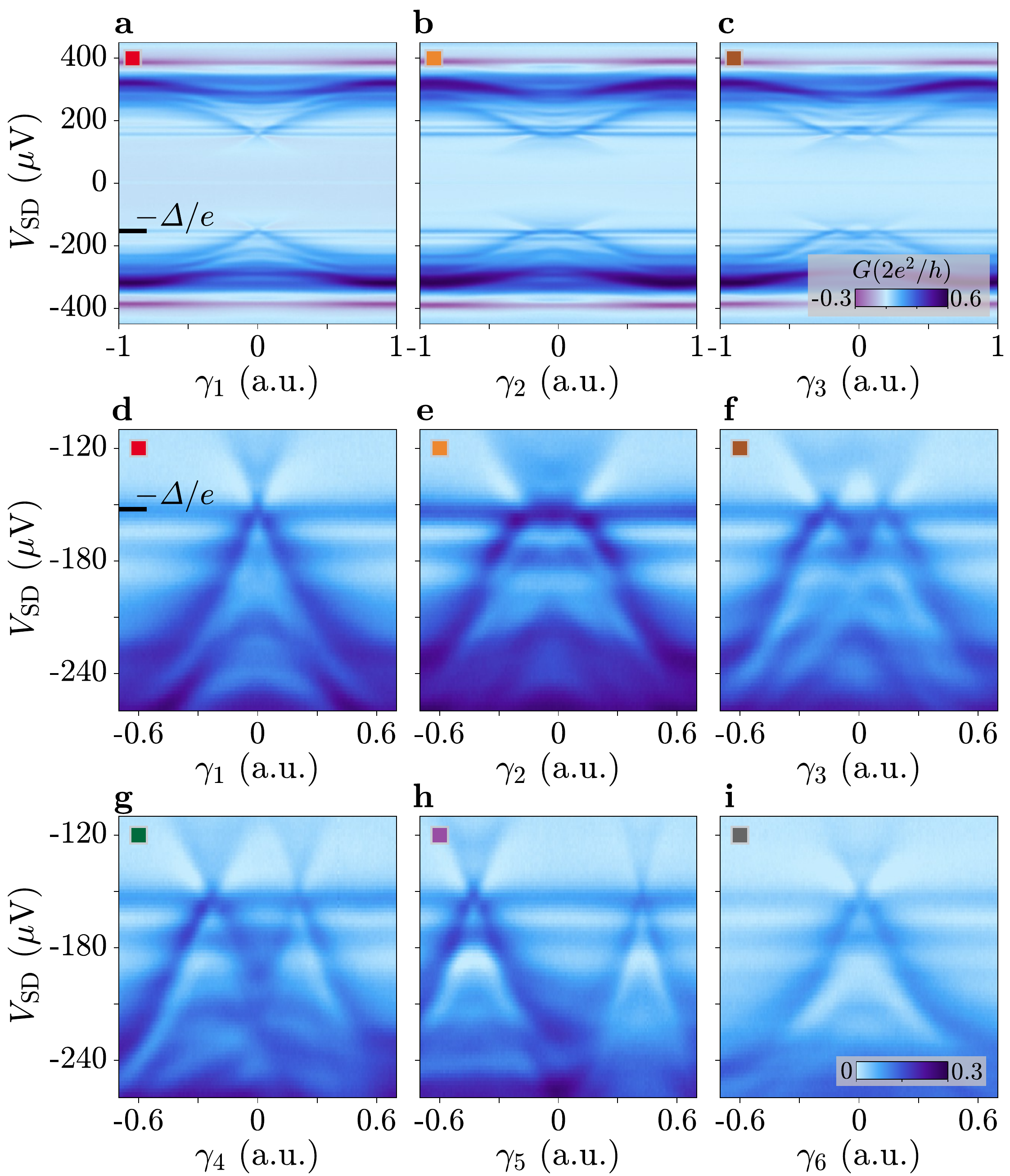}
	\caption{\textbf{Tunnelling conductance spectra along phase space linecuts.} 
		\textbf{a}--\textbf{c}, Differential tunnelling conductance $G$ measured as a function of voltage bias $\Vsd$ along the linecuts $\gamma_i$ (coloured arrows in Fig.~\ref{fig1}d), with $\Vj = 0$. The lower edge of the transport gap $-\mathit{\Delta}/e$, due to the superconducting tunnelling probe, is indicated by the black marker.
		\textbf{d}--\textbf{h}, As \textbf{a}--\textbf{c}, but plotted over restricted ranges of $\Vsd$ and $\gamma_i$.
		\textbf{i}, As \textbf{d}--\textbf{h}, but along linecut $\gamma_6$ (defined in Fig.~\ref{fig1}e), for ${\Vj=-1.5~\mathrm{V}}$. The colourbar in \textbf{i} applies to \textbf{d}--\textbf{i}.}
	\label{fig2}
\end{figure}

\subsection*{Andreev dispersion along phase space linecuts}
With the phase space overview acquired at constant voltage bias, we select cut lines $\gamma_\mathrm{1-6}$ (arrows in Fig.~\ref{fig1}d,e) along which the tunnelling conductance is measured as a function of $\Vsd$. The full datasets along  $\gamma_1$, $\gamma_2$ and $\gamma_3$ are displayed in Fig.~\ref{fig2}a--c. Each shows a transport gap $2 \mathit{\Delta} / e \approx 310~\mu \mathrm{V}$ ($e$ is the elementary charge), consistent with a superconducting gap of the Al probe $\mathit{\Delta} \approx 155~\mu \mathrm{eV}$, and an electron--hole-symmetric spectrum revealing phase-dependent ABSs. The presence of regions with finite conductance at $e|\Vsd| \lesssim \mathit{\Delta}$ is ascribed to broadened features in the DOS of the superconducting probe for energies $|E| \sim \mathit{\Delta}$. This could be due to a combination of quasiparticle-lifetime broadening \cite{Dynes1978} and additional subgap bound states forming between probe and 3TJJ.
On either side of the spectrum, we notice two differential conductance peaks at $\Vsd = \pm 155~ \mu \mathrm{V}$ and $\pm 175~\mu \mathrm{V}$, whose position in bias does not vary appreciably with $\gamma_i$. The first is attributed to the multiple Andreev reflection peak at $\pm \mathit{\Delta} / e$, while the second, specific to this device, might be related to mesoscopic defects in the tunnelling probe or to a region in the device with a larger superconducting gap. These peaks provide a contribution to the measured differential conductance, adding to that of ABS resonances and spectroscopy background. This accounts for the intensity modulation of these peaks depending on $\gamma_i$. Further supported by measurements at different tunings of the probe (see Supplementary Note 4), we do not observe a distortion of the ABS dispersion when the states cross the peaks.
In Fig.~\ref{fig2}d--i, all six linecuts are plotted in restricted $\Vsd$ and $\gamma_i$ ranges for better clarity. In each case, ABSs approach the transport gap edge very closely, which indicates near-unity transmission. We also note that such highly transmissive ABSs intersect $\Vsd = -170~\mu \mathrm{V}$ twice per period, thus accounting for the pairs of resonances in Fig.~\ref{fig1}d,e. Interestingly, when comparing the $\gamma_1$ and $\gamma_2$ linecuts (Fig.~\ref{fig2}d,e), we find anisotropic ABS phase dispersion in the vicinity of $(\pi,\pi)$ phase ($\gamma_\mathrm{1,2}=0$), with a narrow, cusp-like shape versus a broader and flatter peak, respectively. In the spectroscopy measurements performed along $\gamma_\mathrm{3-5}$, i.e., lines parallel to $\gamma_2$ but offset from an intersection point of Fig.~\ref{fig1}d, two distinct highly transmissive ABSs appear. When their separation in phase is small, the states partially mix and an avoided crossing is observed (Fig.~\ref{fig2}f), an effect which is weaker at larger separation (Fig.~\ref{fig2}g), until it is completely suppressed (Fig.~\ref{fig2}h). Finally, linecut $\gamma_6$, where the switch junction is kept in the OFF state, reveals a single highly transmissive ABS (Fig.~\ref{fig2}i).

\begin{figure*}
	\includegraphics[width=\textwidth]{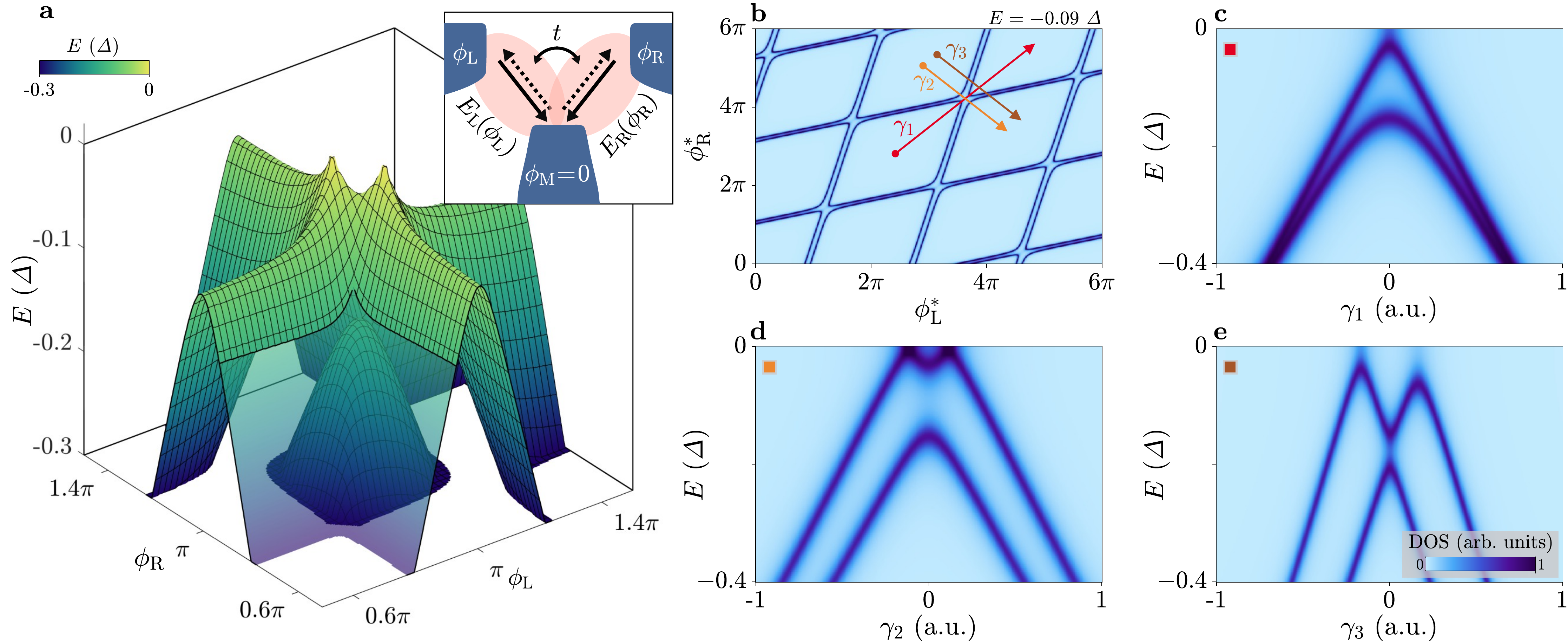}
	\caption{\textbf{Theoretical model of coupled Andreev bound states and Andreev molecule.} 
		\textbf{a}, Simulated Andreev bound state (ABS) energy bands as a function of the superconducting phase differences $\phiL$ and $\phiR$ for negative energies $E \leq 0$. The band structure at positive energies (not shown) is specular due to electron--hole symmetry. Inset: simplified schematic of the model. Three superconducting terminals (blue), with phases $\phiL$, $\phiR$ and $\phiM=0$, are interconnected via two Andreev channels (pink) that are located between the middle lead and either the left or the right lead. The energy $E_\mathrm{L(R)}$ of the left (right) ABS depends on the phase difference $\phi_\mathrm{L(R)}-\phi_\mathrm{M} \equiv \phi_\mathrm{L(R)}$. The two ABSs are coupled to each other, enabling hybridisation of their energy levels, described by the parameter $t$. A full description of the model is provided in the Supplementary Information.
		\textbf{b}, Simulated density of states (DOS) at fixed energy $E = - 0.09 \mathit{\Delta}$ as a function of the cross-coupled superconducting phase differences $\phiL^*$ and $\phiR^*$, defined as linear combinations of $\phiL$ and $\phiR$ to better represent the experimental data (see Supplementary Information for more details).
		\textbf{c}--\textbf{e}, Simulated DOS as a function of energy along the linecuts of the phase space $\gamma_{1-3}$, defined in \textbf{b} (coloured arrows).
	}
	\label{fig3}
\end{figure*}

\subsection*{Andreev molecule model and numerical simulations}
To simulate ABSs arising in a 3TJJ, we study a minimal theoretical model comprising three superconducting terminals with phases $\phiL$, $\phiR$ and $\phiM$, as schematically sketched in the inset of Fig.~\ref{fig3}a. Due to gauge invariance, only two phases are independent, therefore we set $\phiM \equiv 0$. Between lead L (R) and M, we assume a highly transmissive channel, fully described by two coupling parameters to the leads and hosting a spin-degenerate ABS, whose energy $E_\mathrm{L(R)}$ depends on $\phi_\mathrm{L(R)} - \phi_\mathrm{M} \equiv \phi_\mathrm{L(R)}$. The choice of the coupling parameters led to transmissions $T_1 \approx 0.998$ and $T_2 \approx 0.992$ for the left and right ABS, respectively. Coupling between the ABSs is described by the parameter $t$, resulting in a minimal set of five numerical parameters, plus a broadening parameter (more details of the model are explained in Supplementary Note 1).
When the two ABSs are isolated ($t=0$), their energy dispersions solely depend on the transmission of the respective channels \cite{Beenakker1991}. However, for finite coupling $t \neq 0$, hybridisation between the states is enabled in the regions of $(\phiL, \phiR)$ where the isolated energies $E_\mathrm{L}$ and $E_\mathrm{R}$ are comparable, resulting in an Andreev molecule \cite{Pillet2019, Kornich2019}. To study the experimentally relevant case, we assume $t = 1.1 \mathit{\Delta}$, indicating ABSs strongly coupled to each other.
In Fig.~\ref{fig3}a, we display a numerical simulation of the Andreev energy bands as a function of $\phiL$ and $\phiR$ for negative energies (i.e., below the Fermi level $E_\mathrm{F} \equiv 0$), noting that specular bands exist at positive energies owing to electron--hole symmetry. The finite coupling between the ABSs results in hybridised bands with a pronounced splitting. In the region with both phases tuned near to $\pi$, the band closer to zero energy shows a striking anisotropy in the phase space and the dispersion is strongly modified compared to that of high-transmission ABSs in a ballistic two-terminal JJ.
To better compare experimental and numerical results, we calculate the DOS at fixed energy $E = - 0.09 \mathit{\Delta}$ as a function of the cross-coupled phases $\phiL^*$ and $\phiR^*$ (Fig.~\ref{fig3}b), and as a function of energy along the three phase space linecuts $\gamma_{1-3}$ (Fig.~\ref{fig3}c--e). All simulations qualitatively reproduce the key features observed in the measurements of Figs.~\ref{fig1}(d) and \ref{fig2}, and lead to the same order of magnitude for the avoided crossings. In the constant-energy simulation (Fig.~\ref{fig3}b), we confirm the presence of a periodic pattern characterised by avoided crossings and phase shifts of the ABS resonances near the intersection points, where individual ABSs are connected to each other forming closed loops. Further, the spectra presented in Fig.~\ref{fig3}c--e resemble the results shown in Fig.~\ref{fig2}d--f: the ABS dispersion approaching zero energy has a sharp peak along $\gamma_1$ and is broader along $\gamma_2$. The individual ABSs reappear with a significant separation when probed along $\gamma_3$, while maintaining a sizeable avoided crossing.
The experimentally measured phase shifts are larger than those calculated theoretically. The enhanced shift is likely produced by the finite inductive coupling between the loops, which is not included in the numerical model and implies a coupling between the fluxes $\PhiL$ and $\PhiR$. The change of slope of the ABS resonances between Figs.~\ref{fig1}d and \ref{fig1}e is consistent with the phase shift and is related to the same coupling mechanism \cite{vanderPloeg2007, Menke2022}. These effects are discussed in more detail in Supplementary Note 6.

\begin{figure*}
	\includegraphics[width=\textwidth]{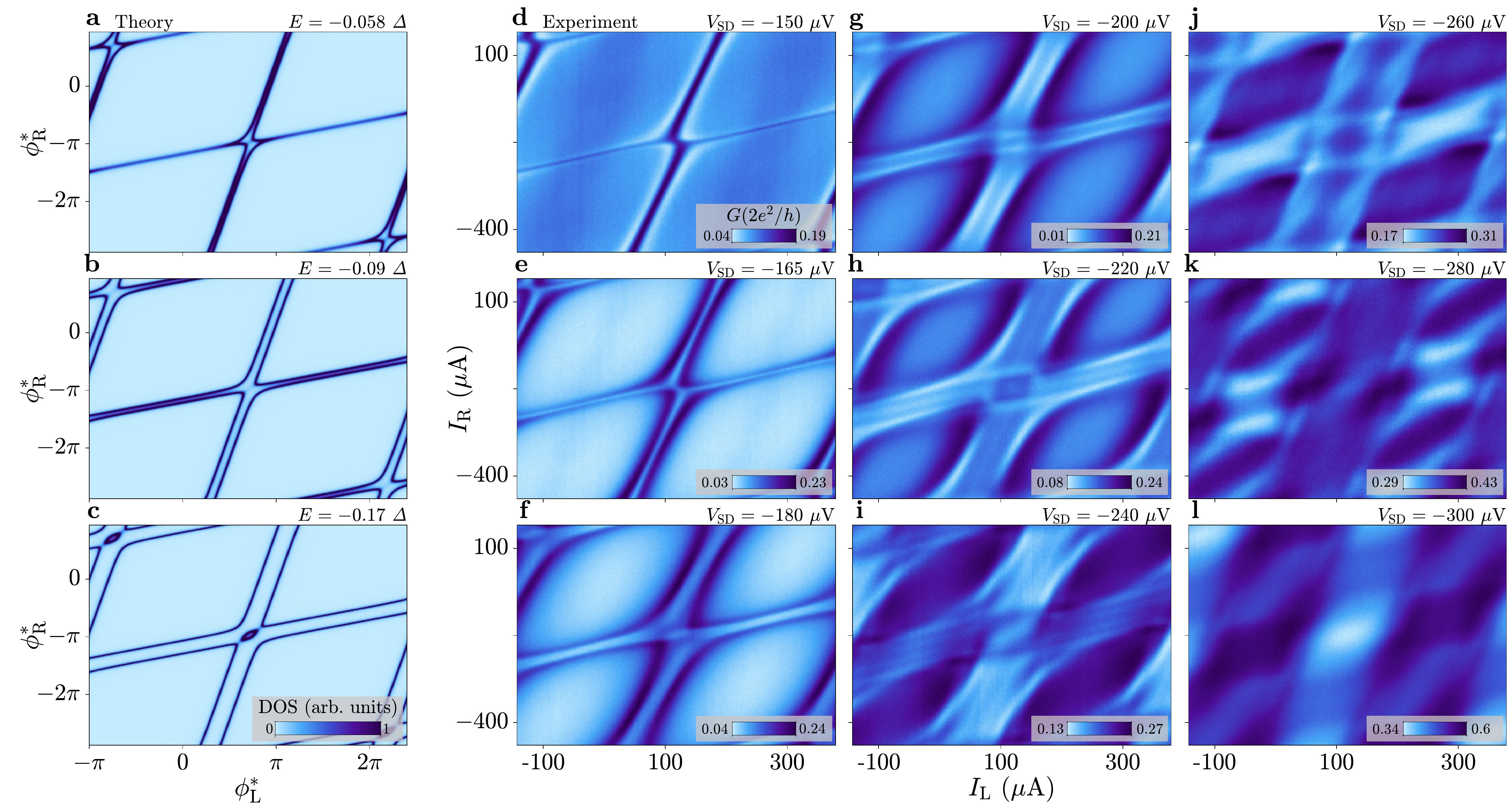}
	\caption{\textbf{Constant-energy planes as a function of the two phases for different energies.}
		\textbf{a}--\textbf{c}, Simulated density of states as a function of the cross-coupled superconducting phase differences $\phiL^*$ and $\phiR^*$ at fixed values of energy $E$.
		\textbf{d}--\textbf{l}, Differential tunnelling conductance $G$ measured as a function of the currents $\Il$ and $\Ir$ injected into the flux-bias lines at fixed values of voltage bias $\Vsd$, for $\Vj=0$.}
	\label{fig4}
\end{figure*}

\subsection*{Tomography of the Andreev band structure}
Another visualisation of the ABS energy bands is provided by combining multiple constant-energy cut planes---each showing the dependence on both superconducting phase differences---to achieve a tomographic representation of the band structure. For this purpose, we measured the tunnelling conductance as a function of $\Il$ and $\Ir$ for several values of $\Vsd$. The outcome is summarised in Fig.~\ref{fig4}, where panels a--c display three theoretically calculated planes in the low-energy spectrum and d--l report experiments for nine values of $\Vsd$. At $\Vsd \approx - \mathit{\Delta} / e $ (i.e., near zero energy), where the ABSs are probed in proximity of their maximum, we observe single resonances (Fig.~\ref{fig4}a,d), which split into pairs at more negative bias (Fig.~\ref{fig4}b,e). We confirm the presence of the features described for Fig.~\ref{fig1}d and Fig.~\ref{fig3}b. Notably, in Fig.~\ref{fig4}c,f resonances related to different states are connected to each other by arcs enclosing an oval region. At even more negative bias, additional resonances arise, compatible with low-transmission ABSs appearing in the spectrum from $\Vsd \approx - 200 ~\mu \mathrm{V}$ (see Fig.~\ref{fig2}). Similar to the high-transmission ABSs discussed previously, low-transmission states first occur as single lines (Fig.~\ref{fig4}g) and then split into pairs (Fig.~\ref{fig4}h,i). Several additional modes emerge at larger $|\Vsd|$, making it difficult to resolve individual states while approaching the continuum (Fig.~\ref{fig4}j--l).
We remark that the theoretical model assumes only two high-transmission modes and does not include any additional states with lower transmission, unlike the experimentally measured spectrum. Therefore, a direct comparison of the spectrum at high $|E|$ is beyond the scope of the model.

Our main experimental findings, including avoided crossings and phase shifts in the constant-bias planes of Figs.~\ref{fig1}d and \ref{fig4}, as well as the anisotropy highlighted in Figs.~\ref{fig2}d,e and \ref{fig3}c,d, were qualitatively reproduced on a second device (see Supplementary Note 3). These measurements suggest the generality of the phenomena observed.

\section*{Discussion}
Supported by our theoretical model of coupled ABSs, we interpret the experimental results summarised in Figs.~\ref{fig1}, \ref{fig2} and \ref{fig4} as evidence of coupling and hybridisation between two highly transmissive ABSs in the normal region of the 3TJJ. Thus, our devices constitute an implementation of an Andreev molecule, comparable to Refs.~\cite{Pillet2019, Kornich2019}. Avoided crossings in both the phase space and the energy spectrum, together with the anisotropic ABS dispersion motivate our interpretation.
Unlike previous experimental studies of Andreev molecules in two-terminal geometries \cite{Su2017, Kurtossy2021}, this realisation is in an open system and is a direct manifestation of the phase-controlled, multidimensional Andreev band structure.

Our system can be considered the magnetic dual of a double quantum dot \cite{Ihn2010}, where electric fields controlled by gate voltages, capacitance and charge on the dots (quantised in units of $e$) are substituted by magnetic fields controlled by currents in flux-bias lines, inductance and magnetic flux threading superconducting loops (quantised in units of the superconducting flux quantum $\mathit{\Phi}_0 = h/2e$, with $h$ the Planck constant).
Both in a double quantum dot and in the present Andreev molecule, overlapping wave functions of two discrete and localised states, coupling in a middle region, result in avoided crossings between their otherwise degenerate energy levels. %Hybridisation is visible in the 2D charge space (charge stability diagram of a double quantum dot) or in the 2D phase space (Fig.~\ref{fig1}d), respectively. 
In our device structure, two discrete levels, namely high-transmission ABSs, form in the short L--M and R--M junctions (whose minimum length is lithographically $30~\mathrm{nm}$) and are coupled to each other by their close proximity.

Spin-resolved ABSs are not observed in the experiments, well described by a spin degenerate model, despite the presence of spin--orbit coupling in our system. We note that here the spin--orbit length, $l_\mathrm{SO} \sim 150~\mathrm{nm}$ for InAs \cite{Fan2020}, is larger than the separation between pairs of terminals of the 3TJJ, resulting into a relatively weak strength of spin--orbit coupling. Enlarging the size of the 3TJJ would thus be required to resolve spin--orbit splitting of ABSs.

In conclusion, ABSs in hybrid 3TJJs were investigated with tunnelling spectroscopy measurements. Owing to the individual control over two superconducting phase differences, we explored a synthetic Andreev band structure and found signatures of coupling and hybridisation between two highly transmissive ABSs, consistent with their overlap in the 3TJJ region and the formation of an Andreev molecule. In the 2D phase space probed at constant voltage bias, we observed periodic patterns with avoided crossings and phase shifts near the intersections between ABS resonances. We measured the spectrum along selected linecuts of the phase space, finding a strong anisotropy of the ABS band structure and avoided crossings between the states. The experiments are well described by a theoretical model of two coupled ABSs.
Our results provide new insights into the physics of multiterminal devices, establish phase control over the ABS band structure and demonstrate the feasibility of realising exotic Andreev matter. Future studies of multidimensional band structures could focus on phase-engineering spin-resolved Andreev levels \cite{vanHeck2014}, ground state fermion parity transitions \cite{vanHeck2014, Whiticar2021, Bargerbos2022} and topological bands, including Weyl singularities \cite{Yokoyama2015, Riwar2016, Eriksson2017, Meyer2017, Xie2017}.

\textit{Note.} We recently became aware of the unpublished data of Refs.~\cite{Lee2022} and \cite{Matsuo2023}, where three-terminal devices were investigated.

\section*{Methods}
\subsection*{Materials and fabrication}
Devices were fabricated in a III--V heterostructure grown with molecular beam epitaxy techniques on an InP (001) substrate. The stack consisted of a step-graded InAlAs buffer layer covered by an $\mathrm{In_{0.75}Ga_{0.25}As}$/InAs/$\mathrm{In_{0.75}Ga_{0.25}As}$ quantum well and two monolayers of GaAs. The InAs layer, hosting a two-dimensional electron gas (2DEG), was 8 nm thick and buried 13 nm below the surface. On top of the III--V stack, a 15 nm thick Al layer was deposited \textit{in situ} without breaking vacuum. Characterisation of the 2DEG in a gated Hall bar revealed a peak mobility of  $18000~\mathrm{cm^{2}V^{-1}s^{-1}}$ at an electron sheet density of ${8\cdot10^{11}~\mathrm{cm^{-2}}}$. This resulted in an electron mean free path $l_{e}\gtrsim260~\mathrm{nm}$, indicating that both the three-terminal Josephson junction and the two-terminal switch junction were in the ballistic regime.

First, large mesa structures were isolated, suppressing parallel conduction between devices and across the middle regions of the superconducting loops. This was done by selectively etching the Al layer with Transene type D, followed by a second chemical etch to a depth of $\sim 380 ~ \mathrm{nm}$ into the III--V material stack, using a $220:55:3:3$ solution of $\mathrm{H_{2}O:C_{6}H_{8}O_{7}:H_{3}PO_{4}:H_{2}O_{2}}$. Next, Al was defined by wet etching with Transene type D at $50\mathrm{^{\circ}C}$ for $4~\mathrm{s}$.
The dielectric, deposited on the entire chip by atomic layer deposition, consisted of a $3~\mathrm{nm}$ thick layer of $\mathrm{Al_2 O_3}$ and a $15~\mathrm{nm}$ thick layer of $\mathrm{HfO_{2}}$. Gate electrodes and flux-bias lines were defined by evaporation and lift-off. In a first step, $5~\mathrm{nm}$ of Ti and $20~\mathrm{nm}$ of Au were deposited to realise the fine features of the gates; in a second, a stack of Ti/Al/Ti/Au with thicknesses $5~\mathrm{nm}$, $340~\mathrm{nm}$, $5~\mathrm{nm}$ and $100~\mathrm{nm}$ was deposited to connect the mesa structure to the bonding pads and to define the flux-bias lines.

\subsection*{Measurement techniques}
Experiments were performed in a dilution refrigerator with base temperature at the mixing chamber below $10~\mathrm{mK}$. The sample was mounted on a QDevil QBoard sample holder system, without employing any light-tight enclosure. Electrical contacts to the devices, excepts for the flux-bias lines, were provided via a resistive loom with QDevil RF and RC low-pass filters at the mixing chamber stage, and RC low-pass filters integrated on the QBoard sample holder. Currents in the flux-bias lines were injected via a superconducting loom with only QDevil RF filters at the mixing chamber stage. Signals were applied to all gates and flux-bias lines via home-made RC filters at room temperature.
Electrical transport measurements were performed with low-frequency AC lock-in techniques. A fixed AC voltage $\delta \Vsd = 5~\mu \mathrm{V}$ at frequency $211~\mathrm{Hz}$ and a variable DC voltage $\Vsd$ were applied to a contact at the superconducting probe (labelled S in Fig.~\ref{fig1}a). The AC current $\delta I$ and the DC current $I_\mathrm{SD}$ flowing in the grounded terminal D were measured via a current-to-voltage (I--V) converter. By measuring the AC voltage $\delta V$ between terminals S and D in a four-terminal configuration, the differential conductance $G \equiv \delta I / \delta V$ was determined.
The refrigerator was equipped with a vector magnet which, despite not being utilised for the experiments, produced a small magnetic field offset. Hence, arbitrary offsets in the flux-bias line currents $\Il$ and $\Ir$ of $-18~\mu \mathrm{A}$ and $74~\mu \mathrm{A}$ were considered in datasets, in such a manner that the point where $\Il = \Ir = 0$ was at the centre of a diamond-like region in the constant-bias maps.

\section*{Data availability}
The data presented in this study have been deposited in Zenodo [\url{https://zenodo.org/record/8360770}]. Further data that support the findings of this study are available upon request from the corresponding author.

\section*{Code availability}
Computer code used to perform the numerical simulations presented in this work has been deposited in Zenodo [\url{https://zenodo.org/record/8360770}].

\bibliography{Bibliography}

\begin{thebibliography}{10}
\expandafter\ifx\csname url\endcsname\relax
  \def\url#1{\texttt{#1}}\fi
\expandafter\ifx\csname urlprefix\endcsname\relax\def\urlprefix{URL }\fi
\providecommand{\bibinfo}[2]{#2}
\providecommand{\eprint}[2][]{\url{#2}}

\bibitem{Andreev1964}
\bibinfo{author}{Andreev, A.~F.}
\newblock \bibinfo{title}{Thermal conductivity of the intermediate state of
  superconductors}.
\newblock \emph{\bibinfo{journal}{Sov. Phys. JETP}}
  \textbf{\bibinfo{volume}{19}}, \bibinfo{pages}{1228--1231}
  (\bibinfo{year}{1964}).

\bibitem{Beenakker1991}
\bibinfo{author}{Beenakker, C. W.~J.} \& \bibinfo{author}{van Houten, H.}
\newblock \bibinfo{title}{{J}osephson current through a superconducting quantum
  point contact shorter than the coherence length}.
\newblock \emph{\bibinfo{journal}{Phys. Rev. Lett.}}
  \textbf{\bibinfo{volume}{66}}, \bibinfo{pages}{3056--3059}
  (\bibinfo{year}{1991}).

\bibitem{Furusaki1991}
\bibinfo{author}{Furusaki, A.} \& \bibinfo{author}{Tsukada, M.}
\newblock \bibinfo{title}{Current-carrying states in {J}osephson junctions}.
\newblock \emph{\bibinfo{journal}{Phys. Rev. B}} \textbf{\bibinfo{volume}{43}},
  \bibinfo{pages}{10164--10169} (\bibinfo{year}{1991}).

\bibitem{Desposito2001}
\bibinfo{author}{Desp\'osito, M.~A.} \& \bibinfo{author}{Levy~Yeyati, A.}
\newblock \bibinfo{title}{Controlled dephasing of {A}ndreev states in
  superconducting quantum point contacts}.
\newblock \emph{\bibinfo{journal}{Phys. Rev. B}} \textbf{\bibinfo{volume}{64}},
  \bibinfo{pages}{140511} (\bibinfo{year}{2001}).

\bibitem{Zazunov2003}
\bibinfo{author}{Zazunov, A.}, \bibinfo{author}{Shumeiko, V.~S.},
  \bibinfo{author}{Bratus', E.~N.}, \bibinfo{author}{Lantz, J.} \&
  \bibinfo{author}{Wendin, G.}
\newblock \bibinfo{title}{{A}ndreev level qubit}.
\newblock \emph{\bibinfo{journal}{Phys. Rev. Lett.}}
  \textbf{\bibinfo{volume}{90}}, \bibinfo{pages}{087003}
  (\bibinfo{year}{2003}).

\bibitem{Chtchelkatchev2003}
\bibinfo{author}{Chtchelkatchev, N.~M.} \& \bibinfo{author}{Nazarov, Y.~V.}
\newblock \bibinfo{title}{Andreev quantum dots for spin manipulation}.
\newblock \emph{\bibinfo{journal}{Phys. Rev. Lett.}}
  \textbf{\bibinfo{volume}{90}}, \bibinfo{pages}{226806}
  (\bibinfo{year}{2003}).

\bibitem{Padurariu2010}
\bibinfo{author}{Padurariu, C.} \& \bibinfo{author}{Nazarov, Y.~V.}
\newblock \bibinfo{title}{Theoretical proposal for superconducting spin
  qubits}.
\newblock \emph{\bibinfo{journal}{Phys. Rev. B}} \textbf{\bibinfo{volume}{81}},
  \bibinfo{pages}{144519} (\bibinfo{year}{2010}).

\bibitem{Pillet2010}
\bibinfo{author}{Pillet, J.-D.} \emph{et~al.}
\newblock \bibinfo{title}{Andreev bound states in supercurrent-carrying carbon
  nanotubes revealed}.
\newblock \emph{\bibinfo{journal}{Nat. Phys.}} \textbf{\bibinfo{volume}{6}},
  \bibinfo{pages}{965--969} (\bibinfo{year}{2010}).

\bibitem{Chang2013}
\bibinfo{author}{Chang, W.}, \bibinfo{author}{Manucharyan, V.~E.},
  \bibinfo{author}{Jespersen, T.~S.}, \bibinfo{author}{Nyg\aa{}rd, J.} \&
  \bibinfo{author}{Marcus, C.~M.}
\newblock \bibinfo{title}{Tunneling spectroscopy of quasiparticle bound states
  in a spinful {J}osephson junction}.
\newblock \emph{\bibinfo{journal}{Phys. Rev. Lett.}}
  \textbf{\bibinfo{volume}{110}}, \bibinfo{pages}{217005}
  (\bibinfo{year}{2013}).

\bibitem{Bretheau2013a}
\bibinfo{author}{Bretheau, L.}, \bibinfo{author}{Girit, {\c{C}}.~{\"O}.},
  \bibinfo{author}{Pothier, H.}, \bibinfo{author}{Esteve, D.} \&
  \bibinfo{author}{Urbina, C.}
\newblock \bibinfo{title}{Exciting {A}ndreev pairs in a superconducting atomic
  contact}.
\newblock \emph{\bibinfo{journal}{Nature}} \textbf{\bibinfo{volume}{499}},
  \bibinfo{pages}{312--315} (\bibinfo{year}{2013}).

\bibitem{Bretheau2013b}
\bibinfo{author}{Bretheau, L.}, \bibinfo{author}{Girit, {\c{C}}.~{\"O}.},
  \bibinfo{author}{Urbina, C.}, \bibinfo{author}{Esteve, D.} \&
  \bibinfo{author}{Pothier, H.}
\newblock \bibinfo{title}{Supercurrent spectroscopy of {A}ndreev states}.
\newblock \emph{\bibinfo{journal}{Phys. Rev. X}} \textbf{\bibinfo{volume}{3}},
  \bibinfo{pages}{041034} (\bibinfo{year}{2013}).

\bibitem{Bretheau2017}
\bibinfo{author}{Bretheau, L.} \emph{et~al.}
\newblock \bibinfo{title}{Tunnelling spectroscopy of {A}ndreev states
  in graphene}.
\newblock \emph{\bibinfo{journal}{Nat. Phys.}} \textbf{\bibinfo{volume}{13}},
  \bibinfo{pages}{756--760} (\bibinfo{year}{2017}).

\bibitem{VanWoerkom2017}
\bibinfo{author}{van Woerkom, D.~J.} \emph{et~al.}
\newblock \bibinfo{title}{{Microwave spectroscopy of spinful {A}ndreev bound
  states in ballistic semiconductor {J}osephson junctions}}.
\newblock \emph{\bibinfo{journal}{Nat. Phys.}} \textbf{\bibinfo{volume}{13}},
  \bibinfo{pages}{876--881} (\bibinfo{year}{2017}).

\bibitem{Tosi2019}
\bibinfo{author}{Tosi, L.} \emph{et~al.}
\newblock \bibinfo{title}{Spin-orbit splitting of {A}ndreev states revealed by
  microwave spectroscopy}.
\newblock \emph{\bibinfo{journal}{Phys. Rev. X}} \textbf{\bibinfo{volume}{9}},
  \bibinfo{pages}{011010} (\bibinfo{year}{2019}).

\bibitem{Nichele2020}
\bibinfo{author}{Nichele, F.} \emph{et~al.}
\newblock \bibinfo{title}{Relating {A}ndreev bound states and supercurrents in
  hybrid {J}osephson junctions}.
\newblock \emph{\bibinfo{journal}{Phys. Rev. Lett.}}
  \textbf{\bibinfo{volume}{124}}, \bibinfo{pages}{226801}
  (\bibinfo{year}{2020}).

\bibitem{Janvier2015}
\bibinfo{author}{Janvier, C.} \emph{et~al.}
\newblock \bibinfo{title}{Coherent manipulation of {A}ndreev states in
  superconducting atomic contacts}.
\newblock \emph{\bibinfo{journal}{Science}} \textbf{\bibinfo{volume}{349}},
  \bibinfo{pages}{1199--1202} (\bibinfo{year}{2015}).

\bibitem{Hays2018}
\bibinfo{author}{Hays, M.} \emph{et~al.}
\newblock \bibinfo{title}{Direct microwave measurement of {A}ndreev-bound-state
  dynamics in a semiconductor-nanowire {J}osephson junction}.
\newblock \emph{\bibinfo{journal}{Phys. Rev. Lett.}}
  \textbf{\bibinfo{volume}{121}}, \bibinfo{pages}{047001}
  (\bibinfo{year}{2018}).

\bibitem{Hays2021}
\bibinfo{author}{Hays, M.} \emph{et~al.}
\newblock \bibinfo{title}{Coherent manipulation of an {A}ndreev spin qubit}.
\newblock \emph{\bibinfo{journal}{Science}} \textbf{\bibinfo{volume}{373}},
  \bibinfo{pages}{430--433} (\bibinfo{year}{2021}).

\bibitem{PitaVidal2023}
\bibinfo{author}{Pita-Vidal, M.} \emph{et~al.}
\newblock \bibinfo{title}{Direct manipulation of a superconducting spin qubit
  strongly coupled to a transmon qubit}.
\newblock \emph{\bibinfo{journal}{Nat. Phys.}} \textbf{\bibinfo{volume}{19}},
  \bibinfo{pages}{1110–1115} (\bibinfo{year}{2023}).

\bibitem{vanHeck2014}
\bibinfo{author}{van Heck, B.}, \bibinfo{author}{Mi, S.} \&
  \bibinfo{author}{Akhmerov, A.~R.}
\newblock \bibinfo{title}{Single fermion manipulation via superconducting phase
  differences in multiterminal {J}osephson junctions}.
\newblock \emph{\bibinfo{journal}{Phys. Rev. B}} \textbf{\bibinfo{volume}{90}},
  \bibinfo{pages}{155450} (\bibinfo{year}{2014}).

\bibitem{Yokoyama2015}
\bibinfo{author}{Yokoyama, T.} \& \bibinfo{author}{Nazarov, Y.~V.}
\newblock \bibinfo{title}{Singularities in the {A}ndreev spectrum of a
  multiterminal {J}osephson junction}.
\newblock \emph{\bibinfo{journal}{Phys. Rev. B}} \textbf{\bibinfo{volume}{92}},
  \bibinfo{pages}{155437} (\bibinfo{year}{2015}).

\bibitem{Riwar2016}
\bibinfo{author}{Riwar, R.-P.}, \bibinfo{author}{Houzet, M.},
  \bibinfo{author}{Meyer, J.~S.} \& \bibinfo{author}{Nazarov, Y.~V.}
\newblock \bibinfo{title}{Multi-terminal {J}osephson junctions as topological
  matter}.
\newblock \emph{\bibinfo{journal}{Nat. Commun.}} \textbf{\bibinfo{volume}{7}},
  \bibinfo{pages}{11167} (\bibinfo{year}{2016}).

\bibitem{Eriksson2017}
\bibinfo{author}{Eriksson, E.}, \bibinfo{author}{Riwar, R.-P.},
  \bibinfo{author}{Houzet, M.}, \bibinfo{author}{Meyer, J.~S.} \&
  \bibinfo{author}{Nazarov, Y.~V.}
\newblock \bibinfo{title}{Topological transconductance quantization in a
  four-terminal {J}osephson junction}.
\newblock \emph{\bibinfo{journal}{Phys. Rev. B}} \textbf{\bibinfo{volume}{95}},
  \bibinfo{pages}{075417} (\bibinfo{year}{2017}).

\bibitem{Meyer2017}
\bibinfo{author}{Meyer, J.~S.} \& \bibinfo{author}{Houzet, M.}
\newblock \bibinfo{title}{Nontrivial {C}hern numbers in three-terminal
  {J}osephson junctions}.
\newblock \emph{\bibinfo{journal}{Phys. Rev. Lett.}}
  \textbf{\bibinfo{volume}{119}}, \bibinfo{pages}{136807}
  (\bibinfo{year}{2017}).

\bibitem{Xie2017}
\bibinfo{author}{Xie, H.-Y.}, \bibinfo{author}{Vavilov, M.~G.} \&
  \bibinfo{author}{Levchenko, A.}
\newblock \bibinfo{title}{Topological {A}ndreev bands in three-terminal
  {J}osephson junctions}.
\newblock \emph{\bibinfo{journal}{Phys. Rev. B}} \textbf{\bibinfo{volume}{96}},
  \bibinfo{pages}{161406} (\bibinfo{year}{2017}).

\bibitem{Deb2018}
\bibinfo{author}{Deb, O.}, \bibinfo{author}{Sengupta, K.} \&
  \bibinfo{author}{Sen, D.}
\newblock \bibinfo{title}{Josephson junctions of multiple superconducting
  wires}.
\newblock \emph{\bibinfo{journal}{Phys. Rev. B}} \textbf{\bibinfo{volume}{97}},
  \bibinfo{pages}{174518} (\bibinfo{year}{2018}).

\bibitem{Xie2019}
\bibinfo{author}{Xie, H.-Y.} \& \bibinfo{author}{Levchenko, A.}
\newblock \bibinfo{title}{Topological supercurrents interaction and
  fluctuations in the multiterminal {J}osephson effect}.
\newblock \emph{\bibinfo{journal}{Phys. Rev. B}} \textbf{\bibinfo{volume}{99}},
  \bibinfo{pages}{094519} (\bibinfo{year}{2019}).

\bibitem{Repin2019}
\bibinfo{author}{Repin, E.~V.}, \bibinfo{author}{Chen, Y.} \&
  \bibinfo{author}{Nazarov, Y.~V.}
\newblock \bibinfo{title}{Topological properties of multiterminal
  superconducting nanostructures: {E}ffect of a continuous spectrum}.
\newblock \emph{\bibinfo{journal}{Phys. Rev. B}} \textbf{\bibinfo{volume}{99}},
  \bibinfo{pages}{165414} (\bibinfo{year}{2019}).

\bibitem{Gavensky2019}
\bibinfo{author}{Peralta~Gavensky, L.}, \bibinfo{author}{Usaj, G.} \&
  \bibinfo{author}{Balseiro, C.~A.}
\newblock \bibinfo{title}{Topological phase diagram of a three-terminal
  {J}osephson junction: {F}rom the conventional to the {M}ajorana regime}.
\newblock \emph{\bibinfo{journal}{Phys. Rev. B}}
  \textbf{\bibinfo{volume}{100}}, \bibinfo{pages}{014514}
  (\bibinfo{year}{2019}).

\bibitem{Houzet2019}
\bibinfo{author}{Houzet, M.} \& \bibinfo{author}{Meyer, J.~S.}
\newblock \bibinfo{title}{Majorana-{W}eyl crossings in topological
  multiterminal junctions}.
\newblock \emph{\bibinfo{journal}{Phys. Rev. B}}
  \textbf{\bibinfo{volume}{100}}, \bibinfo{pages}{014521}
  (\bibinfo{year}{2019}).

\bibitem{Klees2020}
\bibinfo{author}{Klees, R.~L.}, \bibinfo{author}{Rastelli, G.},
  \bibinfo{author}{Cuevas, J.~C.} \& \bibinfo{author}{Belzig, W.}
\newblock \bibinfo{title}{Microwave spectroscopy reveals the quantum geometric
  tensor of topological {J}osephson matter}.
\newblock \emph{\bibinfo{journal}{Phys. Rev. Lett.}}
  \textbf{\bibinfo{volume}{124}}, \bibinfo{pages}{197002}
  (\bibinfo{year}{2020}).

\bibitem{Pillet2019}
\bibinfo{author}{Pillet, J.-D.}, \bibinfo{author}{Benzoni, V.},
  \bibinfo{author}{Griesmar, J.}, \bibinfo{author}{Smirr, J.-L.} \&
  \bibinfo{author}{Girit, {\c{C}}.~O.}
\newblock \bibinfo{title}{Nonlocal {J}osephson effect in {A}ndreev molecules}.
\newblock \emph{\bibinfo{journal}{Nano Lett.}} \textbf{\bibinfo{volume}{19}},
  \bibinfo{pages}{7138--7143} (\bibinfo{year}{2019}).

\bibitem{Kornich2019}
\bibinfo{author}{Kornich, V.}, \bibinfo{author}{Barakov, H.~S.} \&
  \bibinfo{author}{Nazarov, Y.~V.}
\newblock \bibinfo{title}{Fine energy splitting of overlapping {A}ndreev bound
  states in multiterminal superconducting nanostructures}.
\newblock \emph{\bibinfo{journal}{Phys. Rev. Res.}}
  \textbf{\bibinfo{volume}{1}}, \bibinfo{pages}{033004} (\bibinfo{year}{2019}).

\bibitem{Kornich2020}
\bibinfo{author}{Kornich, V.}, \bibinfo{author}{Barakov, H.~S.} \&
  \bibinfo{author}{Nazarov, Y.~V.}
\newblock \bibinfo{title}{Overlapping {A}ndreev states in semiconducting
  nanowires: {C}ompetition of one-dimensional and three-dimensional
  propagation}.
\newblock \emph{\bibinfo{journal}{Phys. Rev. B}}
  \textbf{\bibinfo{volume}{101}}, \bibinfo{pages}{195430}
  (\bibinfo{year}{2020}).

\bibitem{Pillet2020}
\bibinfo{author}{Pillet, J.-D.}, \bibinfo{author}{Benzoni, V.},
  \bibinfo{author}{Griesmar, J.}, \bibinfo{author}{Smirr, J.-L.} \&
  \bibinfo{author}{Girit, {\c{C}}.}
\newblock \bibinfo{title}{Scattering description of {A}ndreev molecules}.
\newblock \emph{\bibinfo{journal}{{SciPost} Phys. Core}}
  \textbf{\bibinfo{volume}{2}}, \bibinfo{pages}{009} (\bibinfo{year}{2020}).

\bibitem{Freyn2011}
\bibinfo{author}{Freyn, A.}, \bibinfo{author}{Dou\ifmmode~\mbox{\c{c}}\else
  \c{c}\fi{}ot, B.}, \bibinfo{author}{Feinberg, D.} \&
  \bibinfo{author}{M\'elin, R.}
\newblock \bibinfo{title}{Production of nonlocal quartets and phase-sensitive
  entanglement in a superconducting beam splitter}.
\newblock \emph{\bibinfo{journal}{Phys. Rev. Lett.}}
  \textbf{\bibinfo{volume}{106}}, \bibinfo{pages}{257005}
  (\bibinfo{year}{2011}).

\bibitem{Jonckheere2013}
\bibinfo{author}{Jonckheere, T.} \emph{et~al.}
\newblock \bibinfo{title}{Multipair dc {J}osephson resonances in a biased
  all-superconducting bijunction}.
\newblock \emph{\bibinfo{journal}{Phys. Rev. B}} \textbf{\bibinfo{volume}{87}},
  \bibinfo{pages}{214501} (\bibinfo{year}{2013}).

\bibitem{Pfeffer2014}
\bibinfo{author}{Pfeffer, A.~H.} \emph{et~al.}
\newblock \bibinfo{title}{Subgap structure in the conductance of a
  three-terminal {J}osephson junction}.
\newblock \emph{\bibinfo{journal}{Phys. Rev. B}} \textbf{\bibinfo{volume}{90}},
  \bibinfo{pages}{075401} (\bibinfo{year}{2014}).

\bibitem{Cohen2018}
\bibinfo{author}{Cohen, Y.} \emph{et~al.}
\newblock \bibinfo{title}{Nonlocal supercurrent of quartets in a three-terminal
  {J}osephson junction}.
\newblock \emph{\bibinfo{journal}{Proc. Natl. Acad. Sci. U.S.A.}}
  \textbf{\bibinfo{volume}{115}}, \bibinfo{pages}{6991--6994}
  (\bibinfo{year}{2018}).

\bibitem{Huang2022}
\bibinfo{author}{Huang, K.-F.} \emph{et~al.}
\newblock \bibinfo{title}{Evidence for 4e charge of {C}ooper quartets in a
  biased multi-terminal graphene-based {J}osephson junction}.
\newblock \emph{\bibinfo{journal}{Nat. Commun.}} \textbf{\bibinfo{volume}{13}},
  \bibinfo{pages}{3032} (\bibinfo{year}{2022}).

\bibitem{Draelos2019}
\bibinfo{author}{Draelos, A.~W.} \emph{et~al.}
\newblock \bibinfo{title}{Supercurrent flow in multiterminal graphene
  {J}osephson junctions}.
\newblock \emph{\bibinfo{journal}{Nano Lett.}} \textbf{\bibinfo{volume}{19}},
  \bibinfo{pages}{1039--1043} (\bibinfo{year}{2019}).

\bibitem{Graziano2020}
\bibinfo{author}{Graziano, G.~V.}, \bibinfo{author}{Lee, J.~S.},
  \bibinfo{author}{Pendharkar, M.}, \bibinfo{author}{Palmstr\o{}m, C.~J.} \&
  \bibinfo{author}{Pribiag, V.~S.}
\newblock \bibinfo{title}{Transport studies in a gate-tunable three-terminal
  {J}osephson junction}.
\newblock \emph{\bibinfo{journal}{Phys. Rev. B}}
  \textbf{\bibinfo{volume}{101}}, \bibinfo{pages}{054510}
  (\bibinfo{year}{2020}).

\bibitem{Pankratova2020}
\bibinfo{author}{Pankratova, N.} \emph{et~al.}
\newblock \bibinfo{title}{Multiterminal {J}osephson effect}.
\newblock \emph{\bibinfo{journal}{Phys. Rev. X}} \textbf{\bibinfo{volume}{10}},
  \bibinfo{pages}{031051} (\bibinfo{year}{2020}).

\bibitem{Matsuo2022}
\bibinfo{author}{Matsuo, S.} \emph{et~al.}
\newblock \bibinfo{title}{Observation of nonlocal {J}osephson effect on double
  {InAs} nanowires}.
\newblock \emph{\bibinfo{journal}{Commun. Phys.}} \textbf{\bibinfo{volume}{5}},
  \bibinfo{pages}{221} (\bibinfo{year}{2022}).

\bibitem{Arnault2021}
\bibinfo{author}{Arnault, E.~G.} \emph{et~al.}
\newblock \bibinfo{title}{Multiterminal inverse {AC} {J}osephson effect}.
\newblock \emph{\bibinfo{journal}{Nano Lett.}} \textbf{\bibinfo{volume}{21}},
  \bibinfo{pages}{9668--9674} (\bibinfo{year}{2021}).

\bibitem{Strambini2016}
\bibinfo{author}{Strambini, E.} \emph{et~al.}
\newblock \bibinfo{title}{The $\omega$-{SQUIPT} as a tool to phase-engineer
  {J}osephson topological materials}.
\newblock \emph{\bibinfo{journal}{Nat. Nanotechnol.}}
  \textbf{\bibinfo{volume}{11}}, \bibinfo{pages}{1055--1059}
  (\bibinfo{year}{2016}).

\bibitem{Shabani2016}
\bibinfo{author}{Shabani, J.} \emph{et~al.}
\newblock \bibinfo{title}{Two-dimensional epitaxial
  superconductor-semiconductor heterostructures: A platform for topological
  superconducting networks}.
\newblock \emph{\bibinfo{journal}{Phys. Rev. B}} \textbf{\bibinfo{volume}{93}},
  \bibinfo{pages}{155402} (\bibinfo{year}{2016}).

\bibitem{Cheah2023}
\bibinfo{author}{Cheah, E.} \emph{et~al.}
\newblock \bibinfo{title}{Control over epitaxy and the role of the {InAs/Al}
  interface in hybrid two-dimensional electron gas systems}.
\newblock \emph{\bibinfo{journal}{Phys. Rev. Mater.}}
  \textbf{\bibinfo{volume}{7}}, \bibinfo{pages}{073403} (\bibinfo{year}{2023}).

\bibitem{Dynes1978}
\bibinfo{author}{Dynes, R.~C.}, \bibinfo{author}{Narayanamurti, V.} \&
  \bibinfo{author}{Garno, J.~P.}
\newblock \bibinfo{title}{Direct measurement of quasiparticle-lifetime
  broadening in a strong-coupled superconductor}.
\newblock \emph{\bibinfo{journal}{Phys. Rev. Lett.}}
  \textbf{\bibinfo{volume}{41}}, \bibinfo{pages}{1509--1512}
  (\bibinfo{year}{1978}).

\bibitem{vanderPloeg2007}
\bibinfo{author}{van~der Ploeg, S. H.~W.} \emph{et~al.}
\newblock \bibinfo{title}{Controllable coupling of superconducting flux
  qubits}.
\newblock \emph{\bibinfo{journal}{Phys. Rev. Lett.}}
  \textbf{\bibinfo{volume}{98}}, \bibinfo{pages}{057004}
  (\bibinfo{year}{2007}).

\bibitem{Menke2022}
\bibinfo{author}{Menke, T.} \emph{et~al.}
\newblock \bibinfo{title}{Demonstration of tunable three-body interactions
  between superconducting qubits}.
\newblock \emph{\bibinfo{journal}{Phys. Rev. Lett.}}
  \textbf{\bibinfo{volume}{129}}, \bibinfo{pages}{220501}
  (\bibinfo{year}{2022}).

\bibitem{Su2017}
\bibinfo{author}{Su, Z.} \emph{et~al.}
\newblock \bibinfo{title}{Andreev molecules in semiconductor nanowire double
  quantum dots}.
\newblock \emph{\bibinfo{journal}{Nat. Commun.}} \textbf{\bibinfo{volume}{8}},
  \bibinfo{pages}{585} (\bibinfo{year}{2017}).

\bibitem{Kurtossy2021}
\bibinfo{author}{K\"{u}rt\"{o}ssy, O.} \emph{et~al.}
\newblock \bibinfo{title}{Andreev molecule in parallel {InAs} nanowires}.
\newblock \emph{\bibinfo{journal}{Nano Lett.}} \textbf{\bibinfo{volume}{21}},
  \bibinfo{pages}{7929--7937} (\bibinfo{year}{2021}).

\bibitem{Ihn2010}
\bibinfo{author}{Ihn, T.}
\newblock \emph{\bibinfo{title}{Semiconductor Nanostructures: Quantum States
  and Electronic Transport}} (\bibinfo{publisher}{Oxford University Press},
  \bibinfo{year}{2010}), \bibinfo{edition}{2} edn.

\bibitem{Fan2020}
\bibinfo{author}{Fan, F.}, \bibinfo{author}{Chen, Y.}, \bibinfo{author}{Pan,
  D.}, \bibinfo{author}{Zhao, J.} \& \bibinfo{author}{Xu, H.~Q.}
\newblock \bibinfo{title}{Measurements of spin{\textendash}orbit interaction in
  epitaxially grown {InAs} nanosheets}.
\newblock \emph{\bibinfo{journal}{Appl. Phys. Lett.}}
  \textbf{\bibinfo{volume}{117}}, \bibinfo{pages}{132101}
  (\bibinfo{year}{2020}).

\bibitem{Whiticar2021}
\bibinfo{author}{Whiticar, A.~M.} \emph{et~al.}
\newblock \bibinfo{title}{Zeeman-driven parity transitions in an {A}ndreev
  quantum dot}.
\newblock \emph{\bibinfo{journal}{Phys. Rev. B}}
  \textbf{\bibinfo{volume}{103}}, \bibinfo{pages}{245308}
  (\bibinfo{year}{2021}).

\bibitem{Bargerbos2022}
\bibinfo{author}{Bargerbos, A.} \emph{et~al.}
\newblock \bibinfo{title}{Singlet-doublet transitions of a quantum dot
  {J}osephson junction detected in a transmon circuit}.
\newblock \emph{\bibinfo{journal}{PRX Quantum}} \textbf{\bibinfo{volume}{3}},
  \bibinfo{pages}{030311} (\bibinfo{year}{2022}).

\bibitem{Lee2022}
\bibinfo{author}{Lee, H.}
\newblock \emph{\bibinfo{title}{Supercurrent and {A}ndreev bound states in
  multi-terminal {J}osephson junctions}}.
\newblock Ph.D. thesis, \bibinfo{school}{University of Maryland}
  (\bibinfo{year}{2022}).

\bibitem{Matsuo2023}
\bibinfo{author}{Matsuo, S.} \emph{et~al.}
\newblock \bibinfo{title}{Phase-dependent {A}ndreev molecules and
  superconducting gap closing in coherently coupled {J}osephson junctions}.
\newblock \bibinfo{howpublished}{arXiv:2303.10540} (\bibinfo{year}{2023}).

\bibitem{Beenakker1997}
\bibinfo{author}{Beenakker, C. W.~J.}
\newblock \bibinfo{title}{Random-matrix theory of quantum transport}.
\newblock \emph{\bibinfo{journal}{Rev. Mod. Phys.}}
  \textbf{\bibinfo{volume}{69}}, \bibinfo{pages}{731--808}
  (\bibinfo{year}{1997}).

\bibitem{Kurilovich2021}
\bibinfo{author}{Kurilovich, P.~D.}, \bibinfo{author}{Kurilovich, V.~D.},
  \bibinfo{author}{Fatemi, V.}, \bibinfo{author}{Devoret, M.~H.} \&
  \bibinfo{author}{Glazman, L.~I.}
\newblock \bibinfo{title}{Microwave response of an {A}ndreev bound state}.
\newblock \emph{\bibinfo{journal}{Phys. Rev. B}}
  \textbf{\bibinfo{volume}{104}}, \bibinfo{pages}{174517}
  (\bibinfo{year}{2021}).

\bibitem{Tinkham2004}
\bibinfo{author}{Tinkham, M.}
\newblock \emph{\bibinfo{title}{Introduction to Superconductivity}}
  (\bibinfo{publisher}{Dover Publications}, \bibinfo{year}{2004}),
  \bibinfo{edition}{2} edn.

\bibitem{Kjaergaard2016}
\bibinfo{author}{Kjaergaard, M.} \emph{et~al.}
\newblock \bibinfo{title}{{Quantized conductance doubling and hard gap in a
  two-dimensional semiconductor–superconductor heterostructure}}.
\newblock \emph{\bibinfo{journal}{Nat. Commun.}} \textbf{\bibinfo{volume}{7}},
  \bibinfo{pages}{12841} (\bibinfo{year}{2016}).

\bibitem{Note_Mgeom}
\bibinfo{note}{$M_\mathrm{geom}$ was calculated analytically as the mutual
  inductance between two rectangular loops of width $a=3.5~\mu \text{m}$ and
  height $b=25~\mu \text{m}$ separated along the direction of their width by
  the distance $c = 0.5~\mu \text{m}$ (i.e., half the width of the Al strips
  constituting the loops in our devices), using $M_\mathrm{geom} = \frac{\mu_0
  b}{2 \pi} \ln{\frac{\left( a + c \right)^2}{c \left( 2a+c \right)}}$. Here,
  $\mu_0$ is the vacuum magnetic permeability.}

\bibitem{TerHaar2005}
\bibinfo{author}{Ter~Haar, A. C.~J.}
\newblock \emph{\bibinfo{title}{Single and coupled {J}osephson junction quantum
  bits}}.
\newblock Ph.D. thesis, \bibinfo{school}{TU Delft} (\bibinfo{year}{2005}).

\bibitem{Paauw2009}
\bibinfo{author}{Paauw, F.~G.}
\newblock \emph{\bibinfo{title}{Superconducting flux qubits: {Q}uantum chains
  and tunable qubits}}.
\newblock Ph.D. thesis, \bibinfo{school}{TU Delft} (\bibinfo{year}{2009}).

\bibitem{Annunziata2010}
\bibinfo{author}{Annunziata, A.~J.} \emph{et~al.}
\newblock \bibinfo{title}{Tunable superconducting nanoinductors}.
\newblock \emph{\bibinfo{journal}{Nanotechnology}}
  \textbf{\bibinfo{volume}{21}}, \bibinfo{pages}{445202}
  (\bibinfo{year}{2010}).

\end{thebibliography}

\clearpage
\newpage
\onecolumngrid

\newcounter{myc} %define new counter
\renewcommand{\thefigure}{\arabic{myc}}
\renewcommand{\figurename}{Supplementary Fig.}

\section*{Supplementary Information: Phase-engineering the Andreev band structure of a three-terminal Josephson junction}

\section*{Supplementary Note 1: Theory}
In the following, we discuss hybridisation of two Andreev bound states (ABSs) between three superconducting leads. The problem has been investigated extensively in the context of the Andreev molecule \cite{Pillet2019, Kornich2019, Kornich2020, Pillet2020}. %In our study, hybridisation occurring directly through the semiconducting region is enabled---qualitatively similar to one-dimensional electron propagation between ABSs in a semiconducting nanowire discussed in \cite{Kornich2019, Kornich2020}. 
Here, we treat hybridisation empirically with a coupling parameter $\gamma$ introducing an avoided crossing in the spectrum. Assuming relatively weak coupling between the two ABSs, the spectrum resembles the energy levels of the independent states except for the points where they would cross, where hybridisation leads to the anticrossing.
A simple perturbation theory can be applied if the energies $E_\mathrm{L,R}$ are sufficiently larger than $\gamma$:
\begin{equation}
	E_{\pm}=\frac{\El \left(\phiL \right) + \Er \left( \phiR \right)}{2} \pm \sqrt{\left( \frac{\gamma}{2} \right)^{2}+\left[\frac{\El \left(\phiL \right)-\Er \left(\phiR \right)}{2} \right]^{2}},
\end{equation}
where 
\begin{equation}
	E_{\alpha} = -\Deltait \sqrt{1-\tau_{\alpha}^2\sin^{2}\frac{\phi_{\alpha}}{2}}, \;\;\;\;\; \alpha \in \left\{ \mathrm{L},\mathrm{R} \right\}
\end{equation}
is the energy of an independent ABS between two superconducting leads with phase difference $\phi_\alpha$ and energy gap $\Deltait$ in a channel with transmission $\tau_\alpha^2$. We recall that the perturbation theory is applicable for $\left|E_{\alpha}(\phi_{\alpha})\right|\gg |\gamma|$: this condition does not hold if the transmissions of the junctions are large ($1-\tau_\alpha \ll 1$) and the phase differences are close to $(2n+1)\pi$, which is the regime studied in the experiment.
Nevertheless, in such experimentally relevant limit we follow the approach of Refs.~\cite{Kornich2019,Kornich2020} and expand the independent ABS energies around $\phi_\alpha=\pi$:
\begin{equation}
	E_\alpha \approx -\Deltait \sqrt{1-\tau_{\alpha}^2 \left[1-\frac{\left(\phi_{\alpha}-\pi\right)^{2}}{4}\right]} \approx-\frac{\Deltait}{2} \sqrt{\left(\phi_{\alpha}-\pi\right)^{2} + \left( 2 r_{\alpha} \right) ^{2}},
\end{equation}
with $r_{\alpha}^{2}=1-\tau_{\alpha}^{2}\ll1$ the reflection amplitude squared.
In the limit of perfect transmission $r_\mathrm{L}=r_\mathrm{R}=0$, the equations derived in Refs.~\cite{Kornich2019,Kornich2020} for the ballistic regime can be employed:
\begin{equation}
	E_{\pm}\approx-\frac{\Deltait}{2}\sqrt{\delta^{2}+\frac{\left( \vphiL + \vphiR \right)^{2}}{4} + \frac{\left(\vphiL - \vphiR\right)^{2}}{4} \pm\left(\vphiL - \vphiR \right)\sqrt{\delta^{2}+\frac{\left(\vphiL+\vphiR\right)^{2}}{4}}},
\end{equation}
where $\varphi_\alpha =\phi_\alpha-\pi$ and $\delta=\gamma/\Deltait$ is the dimensionless coupling parameter. The formula is applicable for any points close to $(\phiL,\phiR)=\left[\pi(2n+1), \, \pi(2m+1)\right]$%where $\vphiL=\phiL-\pi(2n+1)$, $\vphiR=\phiR-\pi(2m+1)$
, with $n$ and $m$ integers. 
The corresponding constant-energy cut plane of the spectrum at $E = -0.01 \Deltait$ is shown in Supplementary Fig.~\ref{figST1}a as a function of the phase differences. Here, hybridisation of purely ballistic channels introduces avoided crossings along the anti-diagonal direction $\phiL = 2 \pi - \phiR$.

Moreover, we consider small but finite reflection amplitudes $r_\alpha\ll1$ for each channel and use the effective Hamiltonian introduced in \cite{Kornich2019} (by projecting on the low-energy states) to derive the spectrum:
\begin{multline}
	E_{\pm}\approx -\frac{\Deltait}{2\sqrt{2}} \times \\ \times \sqrt{2\delta^{2}+\vphiL^{2}+\left(2 r_\mathrm{L} \right)^{2}+\vphiR^{2}+\left(2 r_\mathrm{R} \right)^{2} \pm \sqrt{\left[\vphiL^{2}+\left(2 r_\mathrm{L} \right)^{2}-\vphiR^{2}-\left(2 r_\mathrm{R} \right)^{2}\right]^{2} + 4\delta^{2}\left[\left(\vphiL-\vphiR\right)^{2}+\left(2 r_\mathrm{L} \right)^{2}+\left(2 r_\mathrm{R} \right)^{2}\right]}}.
\end{multline}
The constant-energy plane at $E = -0.01 \Deltait$ is displayed in Supplementary Fig.~\ref{figST1}b, where we observe that non-zero reflection amplitudes introduce additional avoided crossings along the diagonal direction $\phiL = \phiR$.

In order to go beyond the limiting case accessible analytically, valid for weak coupling between the ABSs, we perform numerical simulations of a comparable system.
Andreev states are modelled as two single-level quantum dots (QDs) coupled to three superconducting terminals. A coupling between the QDs accounts for hybridisation of the ABSs in the three-terminal junction. The model is schematically shown in Supplementary Fig.~\ref{figST2}. For each ABS, the only relevant parameter is the transmission of the corresponding junction \cite{Beenakker1997}, hence we can arbitrarily opt for any microscopic model. The QD model offers a flexible tool to describe a scattering region and is well suited for numerical analyses of the system, even for strong coupling between the ABSs, where the analytical approach cannot be used. The coupling between each QD and the superconducting terminals can be arbitrarily strong, hence this model is also suited to describe ABSs in the open regime, rather than isolated dots weakly coupled to the leads \cite{Kurilovich2021}.

The total Hamiltonian of the system is expressed as:
\begin{align}\label{eq:total_H}
	H = H_{\text{DD}}+\sum_{\alpha=\mathrm{M,L,R}} H_{\text{S},\alpha}+H_{\text{DS}} .
\end{align}
Here, $H_{\text{DD}}$ is the Hamiltonian of the double QD:
\begin{align}
	H_{\text{DD}}=\epsilon_1\sum_{\sigma}d_{1\sigma}^\dagger d_{1\sigma}+\epsilon_2\sum_{\sigma}d_{2\sigma}^\dagger d_{2\sigma} + t\sum_{\sigma}d_{1\sigma}^\dagger d_{2\sigma}+\text{h.c.},
\end{align}
where the creation operator $d_{j\sigma}^\dagger$ of the $j$-th QD, its energy $\epsilon_j$ and the coupling parameter $t$ are defined. The superconducting terminals are described by Bardeen--Cooper--Schriffer Hamiltonians \cite{Tinkham2004}:
\begin{align}\label{intro_eq:BCS}
	H_{\text{S},\alpha}=\sum_{\bm{k}\sigma}\xi_{\bm{k}}c_{\bm{k}\sigma,\alpha}^\dagger c_{\bm{k}\sigma,\alpha} + \sum_{\bm{k}}\left(\Deltait \mathrm{e}^{i\phi_\alpha}c_{\bm{k}\uparrow,\alpha}^\dagger c_{-\bm{k}\downarrow,\alpha}^\dagger +\Deltait \mathrm{e}^{-i\phi_\alpha}c_{-\bm{k}\downarrow,\alpha}c_{\bm{k}\uparrow,\alpha}\right),
\end{align}
where $c_{\bm{k}\sigma,\alpha}^\dagger$  is the creation operator of an electron with momentum $\bm{k}$ and spin $\sigma$ in lead $\alpha \in \left\{ \mathrm{M,L,R} \right\}$, $\Deltait$ is the superconducting gap (assumed to be the same for all leads), $\phi_\alpha$ the superconducting phase of lead $\alpha$ and $\xi_{\bm{k}}$ the normal state dispersion in the leads.
The tunnel coupling between each QD and the superconductors is expressed by the term $H_{\text{DS}}$, having the form:
\begin{align}
	H_{\text{DS}}=v_{\mathrm{M},1}\sum_{\bm{k},\sigma}d_{\mathrm{1}\sigma}^\dagger c_{\bm{k}\sigma,\mathrm{M}} + v_{\mathrm{M},2}\sum_{\bm{k},\sigma}d_{\mathrm{2}\sigma}^\dagger c_{\bm{k}\sigma,\mathrm{M}} + v_{\mathrm{L},1}\sum_{\bm{k},\sigma}d_{\mathrm{1}\sigma}^\dagger c_{\bm{k}\sigma,\mathrm{L}} + v_{\mathrm{R},2}\sum_{\bm{k},\sigma}d_{\mathrm{2}\sigma}^\dagger c_{\bm{k}\sigma,\mathrm{R}} + \text{h.c.} ,
\end{align}
where $v_{\alpha,j}$ denotes the coupling of the $j$-th QD to lead $\alpha$.

Two superconductor--QD--superconductor junctions are identified: between leads L and M through QD\textsubscript{1} and between leads R and M through QD\textsubscript{2}. In the absence of coupling between the QDs ($t=0$), the ABS energy of QD\textsubscript{1(2)} depends only on $\phi_\mathrm{L(R)} - \phiM$.
For finite coupling ($t \neq 0$) the ABSs hybridise, which is the case presented in the Main Text.
We do not consider a direct ballistic channel between terminals L and R due to their larger separation in the device compared to the L--M and R--M junctions. Nevertheless, the energy of a hypothetical ABS in such channel would depend on $\phiL - \phiR$, hence it would be close to the superconducting gap in proximity of the points where $\phiL = \phiR = \pi$. As a consequence, any hybridisation of the additional state with the two included in the model would be suppressed near those regions of phase space.
To evaluate this model in the limit of strong coupling between superconducting leads and QDs, i.e., $\Gammait_{\alpha,j} \equiv \pi N_0 v_{\alpha,j}^2>\Deltait$ (where $N_0$ is the normal density of states in the leads), we calculate the ABS density of states by determining the Green's function of the coupled system with the Dyson equation:
\begin{align}
	\bm{G}_{\text{DD}}=\bm{g}_{\text{DD}}+\bm{g}_{\text{DD}} \, \bm{V}_{\text{DS}} \, \bm{g}_{\text{SS}} \, \bm{V}_{\text{DS}}^\dagger \, \bm{G}_{\text{DD}},
\end{align}
where $\bm{G}_{\text{DD}}$ is the dressed Green's function of the double QD, $\bm{g}_{\text{DD}}$ the unperturbed Green's function of the two QDs, $\bm{V}_\text{DS}$ the coupling between the QDs and the superconducting leads and $\bm{g}_{\text{SS}}=\text{diag}\left(\bm{g}_{\text{SS,M}},\,\bm{g}_{\text{SS,L}},\,\bm{g}_{\text{SS,R}}\right)$ the unperturbed Green's function of the three leads. From this expression, the ABS density of states is calculated as $\rho=-\frac{1}{\pi}\text{Im}\left\{\text{tr}\left(\bm{G}_\text{DD}\right)\right\}$ and the eigenenergies are obtained from the poles of $\bm{G}_{\text{DD}}$, yielding the results shown in Figs.~3 and 4 of the Main Text.
As previously discussed, the only relevant parameters for the hybridised ABSs are the inter-QD coupling $t$ and the transmission amplitudes, which are related to the QD--lead couplings $\Gammait_{\alpha,j}$ via the expressions:
\begin{align}
	T_1 =\frac{4\,\Gammait_{\mathrm{M},1}\,\Gammait_{\mathrm{L},1}}{\left(\Gammait_{\mathrm{M},1}+\Gammait_{\mathrm{L},1}\right)^2}, \;\;\;\;\;
	T_2 =\frac{4\,\Gammait_{\mathrm{M},2}\,\Gammait_{\mathrm{R},2}}{\left(\Gammait_{\mathrm{M},2}+\Gammait_{\mathrm{R},2}\right)^2},
\end{align}
valid in the limit of strong lead--QD coupling. These parameters influence the region of the avoided crossing between the two ABS, as illustrated in Supplementary Fig.~\ref{figST3}. In all simulations, a broadening parameter $\eta$ was assumed in the Green's functions. We observe qualitative agreement between the numerical simulations and the analytical study previously discussed (see Supplementary Fig.~\ref{figST1}).

The parameters used for the simulations in the Main Text were: $t=1.1 \Deltait$, $\Gammait_{\mathrm{L},1}=5.5 \Deltait$, $\Gammait_{\mathrm{M},1}=\Gammait_{\mathrm{R},2}=6 \Deltait$, $\Gammait_{\mathrm{M},2}=5 \Deltait$, corresponding to transmissions $T_1 \approx 0.998$ and $T_2 \approx 0.992$, and broadening ${\eta = 0.02 \Deltait}$.
Moreover,  we recall that the constant-energy planes of Figs.~3b and 4a--c were plotted by introducing a cross-dependence between the phase differences $\phiL-\phiM \equiv \phiL$ and $\phiR-\phiM \equiv \phiR$ to better represent the experimental data, where a cross-dependence between the two flux-bias lines was present (see also discussion in Section \ref{Section_M}). In particular, the cross-coupled phases $\phiL^*$ and $\phiR^*$ were defined as linear combinations of $\phiL$ and $\phiR$ with the transformation:
\begin{align}
	\begin{pmatrix}
		\phiL^* \\
		\phiR^*
	\end{pmatrix}
	=
	\begin{pmatrix}
		a  & b \\
		c  & d
	\end{pmatrix}
	\cdot
	\begin{pmatrix}
		\phiL \\
		\phiR
	\end{pmatrix},
\end{align}
where the coefficients $a = 0.9708$, $b=0.2400$, $c=0.2832$ and $d=1.122$ were used.

\section*{Supplementary Note 2: Gate dependence of differential conductance}

The transmission between the superconducting probe and the three-terminal Josephson junction (3TJJ) was controlled with two gate electrodes, denoted tunnel gates and set to the same voltage $\Vt \equiv \Vtl = \Vtr$, and had a finite dependence on the gates energised by the voltages $\Vp$ and $\Vg$, denoted probe gate and global gate respectively (see Fig.~1a,c). We always kept $\Vp=150~\mathrm{mV}$. The effect of the gate at voltage $\Vj$ (switch gate) on the probe transmission was negligible. The differential conductance $G$ measured as a function of $\Vt$ and $\Vsd$, with $\Vsd$ the DC voltage bias applied to the probe, is shown in Supplementary Fig.~\ref{Sfig1}a,b for $\Vg = 50~\mathrm{mV}$ and $\Vg = -150~\mathrm{mV}$ respectively. In both cases the conductance, hence the probe transmission, decreases with $\Vt$, transitioning from open to tunnelling regime. In the former, we observe a zero-bias conductance peak, corresponding to a remnant of supercurrent between the probe and the superconducting leads, and several peaks at finite bias, related to multiple Andreev reflection (MAR) processes. In the tunnelling regime, where the conductance at large $|\Vsd|$ is substantially lower than the conductance quantum $G_{0}=2e^{2}/h$, a transport gap of $\approx 310~\mu \mathrm{V}$ is present in the spectrum, consistent with a superconducting gap of Al $\mathit{\Delta} \approx 155~\mu \mathrm{eV}$. Pronounced features appearing at the edges of the gap correspond to Andreev bound states (ABSs). As shown in Supplementary Fig.~\ref{Sfig1}b, by further lowering $\Vt$ the probe could be completely pinched off.
We remark that tunnelling spectroscopy in our devices was performed in a superconductor--insulator--superconductor (SIS) configuration, hence the $G(\Vsd)$ traces result from a convolution product between two relatively complex densities of states \cite{Pillet2010, Nichele2020} and do not provide a direct measurement of the gap hardness. Nevertheless, since $G \approx 0$ across a significant voltage range around $\Vsd=0$ (for any flux-line currents $\Il$ and $\Ir$, as seen in Fig.~2 of the Main Text), we consider the gap hardness to be comparable to previous reports \cite{Kjaergaard2016}, where also a density-of-states broadening to what we noted in the Main Text was observed.

The dependence on the global gate is presented in Supplementary Fig.~\ref{Sfig1}c for $\Vt=-1.07~\mathrm{V}$. We find that its main effect on the spectrum is also to change the transmission of the probe, which is progressively reduced for decreasing $\Vg$ until the pinch off is reached. Notably, in both the $\Vt$ and the $\Vg$ dependence, no sharp conductance peaks relatable to resonant transport via spurious quantum dots are observed.

Tunnelling spectroscopy measurements were performed on a second device, fabricated on the same chip of the first and measured in the same cool down. Device 2 was lithographically similar to Device 1, except for the width of the superconducting probe and the shape of the tunnel and probe gates (see Supplementary Fig.~\ref{Sfig1}d). The dependences on the tunnel gate voltage $\Vt$ and on the global gate voltage $\Vg$ are plotted in Supplementary Fig.~\ref{Sfig1}e and f respectively, showing features qualitatively very similar to Device 1.

\section*{Supplementary Note 3: Results for Device 2}

The main experimental results shown in the Main Text for Device 1 were qualitatively reproduced in Device 2, as illustrated in Supplementary Figs.~\ref{figED1} and \ref{figED2}. Here, the tunnel gates were set to $\Vt = -1.395~\mathrm{V}$ and $\Vt = -1.42~\mathrm{V}$ respectively, with the global gate voltage set to $\Vg = -150~\mathrm{mV}$ and the probe gate voltage to $\Vp = 100~\mathrm{mV}$. In these configurations, the probe was in the tunnelling regime and its transmission was comparable to that of Device 1 for the measurements presented in the Main Text. 

In Supplementary Fig.~\ref{figED1}, we show the constant-bias planes (i.e., $G$ as a function of $\Il$ and $\Ir$ at fixed values of $\Vsd$) corresponding to Figs.~1d,e, and 4d--l of the Main Text. When the switch junction is in the ON state (defined by $\Vj = 0$), we confirm the presence of avoided crossings between a $\PhiL$-dependent ABS and a $\PhiR$-dependent ABS, with the resonances associated to one state connecting to those of the other, and phase shifts occurring near the avoided crossings. The 2D pattern is strongly simplified when the switch is OFF ($\Vj = -1.5~\mathrm{V}$, Supplementary Fig.~\ref{figED1}b), as the $\PhiR$-dependent ABS disappears and no sign of hybridization is observed. The band structure tomography is displayed in Supplementary Figs.~\ref{figED1}c--k and is compatible with the results presented in the Main Text (see Fig.~4). 

Further, we select linecuts of the phase space, indicated by the coloured arrows in Supplementary Figs.~\ref{figED1}a,b, along which we perform bias-dependent spectroscopy (Supplementary Fig.~\ref{figED2}, to be compared with Fig.~2). Again, we observe strong dispersion anisotropy when comparing linecut $\gamma_1$ to $\gamma_2$. Incidentally, while we still note a conductance peak at $\Vsd = \pm 155~\mu \mathrm{V}$, whose position in bias does not vary appreciably with $\gamma_i$ and which is attributed to MAR processes, this device does not reveal a second peak at larger $|\Vsd|$. This shows that the peak at $\Vsd = \pm 175~\mu \mathrm{V}$ in Device 1 is a device-specific feature. Since the ABS dispersion of Device 1 is qualitatively reproduced in Device 2, we corroborate that the peak at $\pm 175~\mu \mathrm{V}$ in the former does not interact with the ABSs or affect their main properties.

\section*{Supplementary Note 4: Results for different $\Vg$}

The measurements of Device 1 shown in the Main Text were obtained in a single gate configuration (except for the switch gate voltage, set to $-1.5~\mathrm{V}$ for Figs.~1e, 2i and to $0~\mathrm{V}$ elsewhere). The main results were reproduced in different gate configurations, defined by setting a new value of $\Vg$ and adjusting $\Vt$ to maintain a comparable probe transmission, crucially remaining in the tunnelling regime. In Supplementary Figs.~\ref{Sfig2}--\ref{Sfig7} we show the three cases $\Vg=-66~\mathrm{mV}$, $\Vg=-150~\mathrm{mV}$ and $\Vg=-250~\mathrm{mV}$, as described in the following. Remarkably, all the key features discussed in the Main Text were qualitatively reproduced in these configurations, highlighting the generality of our results and further supporting our conclusions. Incidentally, we did not observe a variation in the transmission of the highly transmissive ABSs as a function of the global gate voltage, namely they approached the edges of the transport gap at $|\Vsd|\approx 155 ~\mu \mathrm{V}$ very closely for any $\Vg$. This was verified in the range $ -300 ~ \mathrm{mV} \leq \Vg \leq 150 ~ \mathrm{mV}$, noting that at $\Vg = -300 ~ \mathrm{mV}$ the sub-gap states were at a limit of visibility but still preserved the same dispersion with phase.

\begin{itemize}
	\item \textbf{$\Vg = -66~\mathrm{mV}$}: Supplementary Figs.~\ref{Sfig2}a,b, \ref{Sfig2}c--k and \ref{Sfig3} correspond to Figs.~1d,e, 4 and 2a--h respectively. $\Vt$ is adjusted to $-1.038~\mathrm{V}$.
	Interestingly, we observe that the two ABSs have very similar visibility in this regime (namely, the resonances associated with each state have similar conductance), in contrast to the regime shown in the Main Text ($\Vg=50~\mathrm{mV}$), where the ABS dispersing with $\PhiL$ had higher conductance. 
	As a possible explanation, we hypothesise that, when $\Vg$ is reduced, the electrostatic potential profile at the tunnelling barrier shifts slightly, in such a manner that its maximum displaces from left to right. At the optimised value $\Vg = -66~\mathrm{mV}$, transport between the probe and either side of the three-terminal region is symmetric, resulting in similar conductance when tunnelling into either ABS.
	Furthermore, in the constant-bias planes of Supplementary Fig.~\ref{Sfig2}, and particularly in panel b, we notice an additional set of resonances with the same dependence on $\Il$ and $\Ir$ as the left ABS. This is attributed to a very small phase modulation of the MAR peak, i.e., the nearly flat line at $\Vsd=-155~\mu \mathrm{V}$ that is visible in Supplementary Fig.~\ref{Sfig3}. These peak, particularly prominent in the present regime, accounts for a conductance background in the constant-bias maps and its modulation translates into a phase-periodic background.
	% OLD: This effect could be ascribed to a slight displacement (from left to right) of the electrostatic potential maximum as the global gate voltage is reduced, thus having more symmetric transport between the probe and either side of the three-terminal region when the global gate voltage is tuned to $-66~\mathrm{mV}$.
	
	\item \textbf{$\Vg = -150~\mathrm{mV}$}: Supplementary Figs.~\ref{Sfig4}a,b, \ref{Sfig4}c--k and \ref{Sfig5} correspond to Figs.~1d,e, 4 and 2 respectively. % maybe mention that linecut of the switch off case was taken at different angle
	$\Vt$ is adjusted to $-1.02~\mathrm{V}$, where the probe transmission is smaller than in the previous regimes: the differential conductance at large $\Vsd$ (near the range limits) is reduced by approximately a factor $2$. This results in a lower signal-to-noise ratio, due to which periodic noise features become visible in the constant-bias measurements (Supplementary Fig.~\ref{Sfig4}), as long as $G$ is relatively small. At the same time, the conductance resonances at $\Vsd = \pm 155~ \mu \mathrm{V}$ and $\Vsd = \pm 175~\mu \mathrm{V}$ in the spectra of Supplementary Fig.~\ref{Sfig5} become substantially less prominent, and so is their modulation as a function of $\gamma_i$. Visibly, the ABS dispersion is not distorted when the states cross the horizontal peaks, remaining qualitatively very similar to that observed in Fig.~2. This provides further support to the absence of an interaction between the peaks at $\Vsd = \pm 155~ \mu \mathrm{V}, \pm 175~\mu \mathrm{V}$ and the ABSs.
	Finally, in this gate configuration, the ABS depending on $\PhiR$ shows the highest conductance, consistent with the argument presented for the case $\Vg=-66~\mathrm{mV}$.
	
	\item \textbf{$\Vg = -250~\mathrm{mV}$}: Supplementary Figs.~\ref{Sfig6}a,b, \ref{Sfig6}c--k and \ref{Sfig7} correspond to Figs.~1d,e, 4 and 2a,b,d,e,i respectively. $\Vt$ is adjusted to $-984~\mathrm{mV}$, where the probe transmission is comparable to the regime $\Vg = -150~\mathrm{mV}$.
\end{itemize}

\section*{Supplementary Note 5: Mutual inductance matrix}\label{Section_M}

As described in the Main Text, our devices feature two flux bias lines where currents $\Il$ and $\Ir$ are injected. Since $\Il$ generates a magnetic field that is stronger over the left superconducting loop than over the right one, it controls mainly the external magnetic flux $\PhiL$ threading the left loop, associated to the superconducting phase difference between the terminals L and M (see Fig.~1a). Similarly, $\Ir$ tunes mostly $\PhiR$, thus the phase difference between R and M, in such a manner that the combination of the two flux lines enables full phase control over a two-dimensional space. Nevertheless, each flux line has also a finite coupling to the opposite loop, thus $\PhiL$ and $\PhiR$ depend on both $\Il$ and $\Ir$ according to the linear relation:
\begin{equation}\label{eqM}
	\begin{pmatrix}
		\PhiL \\
		\PhiR
	\end{pmatrix}
	= \mathbf{M} \cdot 
	\begin{pmatrix}
		\Il \\
		\Ir
	\end{pmatrix}
	=
	\begin{pmatrix}
		M_\mathrm{LL}  & M_\mathrm{LR} \\
		M_\mathrm{RL}  & M_\mathrm{RR}
	\end{pmatrix}
	\cdot
	\begin{pmatrix}
		\Il \\
		\Ir
	\end{pmatrix},
\end{equation}
where $\mathbf{M}$ is the mutual inductance matrix. We calculate $\mathbf{M}$ from the constant-bias conductance measurement of Fig.~1d, plotted again in Supplementary Fig.~\ref{Sfig8}a. First, we observe that the origin $(\Il, \Ir) = (0, 0)$ corresponds to the origin of the flux space $(\PhiL, \PhiR) = (0, 0)$. We associate the centres of the adjacent diamond-like regions to the addition or subtraction of one superconducting flux quantum $\mathit{\Phi}_0 = h/(2e)$ to either $\PhiL$ or $\PhiR$. By evaluating Eq.~\ref{eqM} in two centre points, for example those related to $(\mathit{\Phi}_0, 0)$ and $(0, \mathit{\Phi}_0)$, we write a $4 \times 4$ equation system and determine:
\begin{equation}\label{eqM2}
	\mathbf{M} = 
	\begin{pmatrix}
		6.96 ~ \mathrm{pH}  & -1.43 ~ \mathrm{pH} \\
		-1.69 ~ \mathrm{pH} & 5.79 ~ \mathrm{pH}
	\end{pmatrix}
	.
\end{equation}
The ratio $M_\mathrm{RR}/M_\mathrm{LL} \approx M_\mathrm{LR}/M_\mathrm{RL} \approx 0.8$ approximately corresponds to the ratio between the length of the vertical segments of the right and left flux line.
Knowing $\mathbf{M}$, we can apply the linear transformation of Eq.~\ref{eqM} to convert the $(\Il,\Ir)$ axes to the $(\PhiL,\PhiR)$ axes, as illustrated in Supplementary Fig.~\ref{Sfig8}b for the dataset of panel a.

\section*{Supplementary Note 6: Phase shifts and inductances in the system}

In the constant-bias measurements showing the dependence on $\Il$ and $\Ir$ with the switch junction ON, such as in Figs.~1d and 4d--i, we remarked the presence of phase shifts occurring when resonances associated to different states (i.e., the $\PhiL$- and the $\PhiR$-dispersing ABS) intersect each other. Concomitantly, we noted a slope variation of the $\PhiL$-dispersing ABS depending on the ON/OFF state of the switch junction (yellow dashed line in Fig.~1e).
In this section, we quantify the shifts and relate them to the different inductive contributions in the device. We find that a Josephson-like mutual inductive coupling between the two loops accounts for the presence of both the shifts and the slope difference.

Since a shift is present for both ABSs, it corresponds to a vector $(\Delta \PhiL,~ \Delta \PhiR)$ in the space of the two external magnetic fluxes $\PhiL$ and $\PhiR$. 
For the extraction of this vector, we consider again the dataset of Fig.~4e from the Main Text, shown also in Supplementary Fig.~\ref{Sfig8}c and, upon applying the basis transformation described in Section \ref{Section_M}, in Supplementary Fig.~\ref{Sfig8}d as a function of $\PhiL$ and $\PhiR$. As a guide for the eye, we overlay lines following the dips between pairs of ABS resonances. In panel c, we mark the (vector) shifts $\mathbf{\Delta I}^*_\mathrm{L}$ and $\mathbf{\Delta I}^*_\mathrm{R}$ along the periodicity directions, while in panel d the shifts $\Delta \PhiL$ and $\Delta \PhiR$ are by construction parallel to the axes. These quantities are linked by the following relations:
\begin{equation}
	\begin{pmatrix}
		\Delta \PhiL \\
		0
	\end{pmatrix}
	= \mathbf{M} \cdot \mathbf{\Delta I}^*_\mathrm{L}, \; \; \; \; \;
	\begin{pmatrix}
		0 \\
		\Delta \PhiR
	\end{pmatrix}
	= \mathbf{M} \cdot \mathbf{\Delta I}^*_\mathrm{R}.
\end{equation} 
From the data, we find $\Delta \PhiL \approx \Delta \PhiR \approx 0.092 \cdot \mathit{\Phi}_0 \equiv \Delta  \mathit{\Phi} $, showing that the effect is symmetric for the two ABSs. 

This phase shift is then expressed as $\Delta \mathit{\Phi} = M_\mathrm{cpl} \cdot I_\mathrm{circ}$, namely as the product between a mutual inductance, accounting for the coupling between two fluxes, and a circulating current in either loop.
Neither $M_\mathrm{cpl}$ nor $I_\mathrm{circ}$ can be directly determined from the experimental data (only their product $\Delta \mathit{\Phi}$), therefore we estimate the inductive coupling term considering different origins. The geometric mutual inductance between the two loops of our devices is $M_\mathrm{geom} \approx 7.3 ~ \mathrm{pH}$ \cite{Note_Mgeom}. Since the loops share a strip of epitaxial Al, its inductance, dominated by the kinetic inductance $L_\mathrm{k}$, has to be added to $M_\mathrm{geom}$ \cite{TerHaar2005,Paauw2009}. The kinetic inductance of the strip is estimated as \cite{Annunziata2010}:
\begin{equation}
	L_\mathrm{k} = \frac{l}{w}\frac{h}{2 \pi^2}\frac{R_\square}{\mathit{\Delta}} \approx 44 ~ \mathrm{pH},
\end{equation}
where $l=25~\mu \mathrm{m}$ and $w=1~\mu \mathrm{m}$ are the length and width of the strip, $R_\square \approx 1.5 ~ \Omega$ the normal state resistivity of the heterostructure stack measured in a Hall bar geometry (where the Al film was not removed) and $\mathit{\Delta} \approx 180 ~\mu \mathrm{eV}$ the superconducting gap of Al. The geometric and kinetic contribution lead to a combined mutual inductance $M_\mathrm{loops} \approx 51.3 ~ \mathrm{pH}$. If we assume that $M_\mathrm{cpl} = M_\mathrm{loops}$, we require a circulating current $I_\mathrm{circ} \approx 3.7 ~\mu \mathrm{A}$ for the shift $\Delta \mathit{\Phi} = 0.092 \cdot \mathit{\Phi}_0$. However, this value is much larger than any reasonable estimate of the critical current between each pair of terminals, which constitutes an upper bound to the supercurrent circulating in the loops. As a consequence, $M_\mathrm{cpl}$ must be substantially larger than $M_\mathrm{loops}$ to limit $I_\mathrm{circ}$. 

The only contribution that we have not discussed so far is a Josephson-like coupling due to the 3TJJ.
Based on its geometry, we expect our 3TJJ to have a critical current of the order of $100 ~ \mathrm{nA}$ and a Josephson inductance of a few nH, which would explain the phase shift to a good degree. 
We deduce that, in our devices, the phase shifts derive from a Josephson-like coupling term rather than the geometric and kinetic inductive couplings, in analogy with existing couplers between superconducting qubits containing Josephson elements \cite{vanderPloeg2007, Menke2022}.

This mutual coupling between $\PhiL$ and $\PhiR$ is also responsible for the slope difference depending on the state of the switch junction (see Fig.~1d,e). In the OFF state, no circulating current can flow in the right loop, thus no mutual effect is present. In the $(\Il,\Ir)$ plane (Fig.~1e of the Main Text), the ABS resonance is parallel to the $\PhiR$ axis (indicated in Fig.~1d). When the switch junction is ON, the slope of the ABSs deviates from the directions of $\PhiL$ and $\PhiR$. This is clearly visible on the $(\PhiL,\PhiR)$ axes (Supplementary Fig.~\ref{Sfig8}b,d), as the resonances are not vertical or horizontal. In fact, due to the mutual coupling, the variation of a flux (for example $\PhiL$) from $0$ to $\mathit{\Phi}_0 / 2$ causes a gradual change in the other ($\PhiR$), up to a maximum shift. Upon crossing $\PhiL = \mathit{\Phi}_0 / 2$, the circulating current in the left loop reverses direction, consequently the induced current in the right loop and the $\PhiR$-shift also flip sign. Therefore, the finite ABS slopes on the $(\PhiL,\PhiR)$ axes (corresponding to the slope difference depending on the state of the switch junction) and the phase shifts are consistent. Both result from the finite Josephson-like mutual inductive coupling between the two loops.

\newpage

\section*{Supplementary Figures}

% Figures
\setcounter{myc}{1}
\begin{figure}[h!]
	\includegraphics[width=0.5\columnwidth]{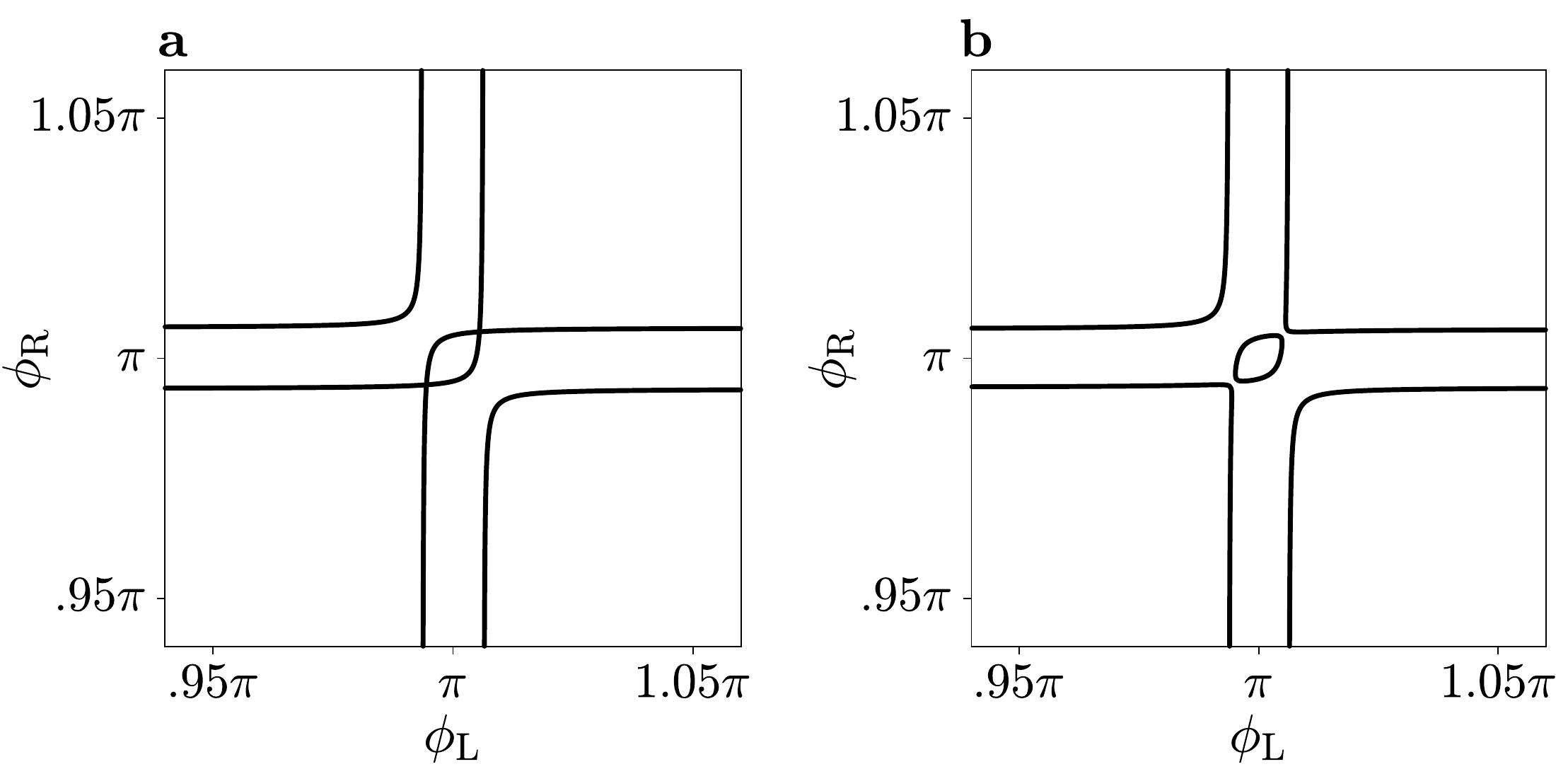}
	\caption{\textbf{Analytical calculations.} 
		\textbf{a},\textbf{b}, Constant-energy planes at $E=-0.01 \Deltait$ as a function of the superconducting phase differences $\phiL$ and $\phiR$ for coupling $\delta=0.01$, in the fully ballistic regime $r_\mathrm{L}=r_\mathrm{R}=0$ (\textbf{a}) and with finite normal reflection amplitudes $r_\mathrm{L}=0.003$, $r_\mathrm{R}=0.002$ (\textbf{b}).}
	\label{figST1}
\end{figure}

\setcounter{myc}{2}
\begin{figure}[h!]
	\includegraphics[width=0.3\columnwidth]{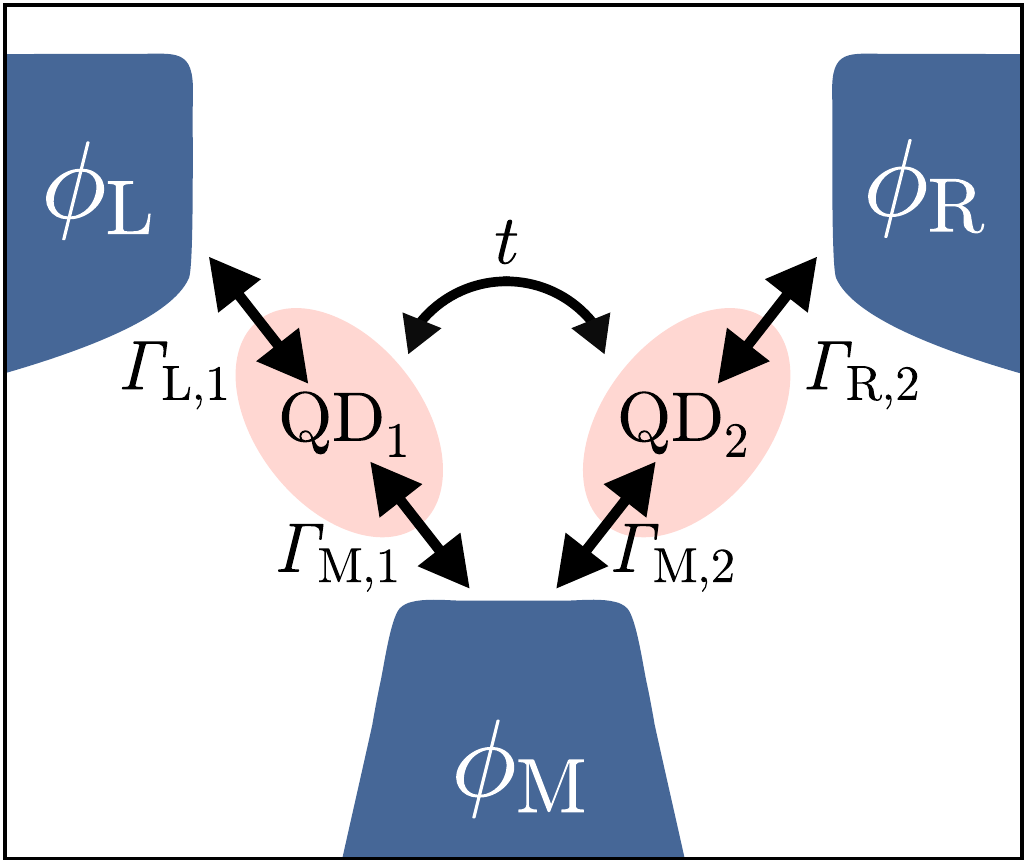}
	\caption{\textbf{Schematic of the numerical model.} 
		Andreev bound states in the three-terminal Josephson junction are modelled as two single-level quantum dots (QDs) coupled to three superconducting leads with phases $\phiL$, $\phiR$ and $\phiM$. QD\textsubscript{1(2)} is coupled to leads L (R) and M, as represented by the parameters $\Gammait_\mathrm{L,1}$ and $\Gammait_\mathrm{M,1}$ ($\Gammait_\mathrm{R,2}$ and $\Gammait_\mathrm{M,2}$) respectively. The parameter $t$ accounts for the coupling between the QDs, enabling their hybridisation.}
	\label{figST2}
\end{figure}

\setcounter{myc}{3}
\begin{figure}[h!]
	\includegraphics[width=0.5\columnwidth]{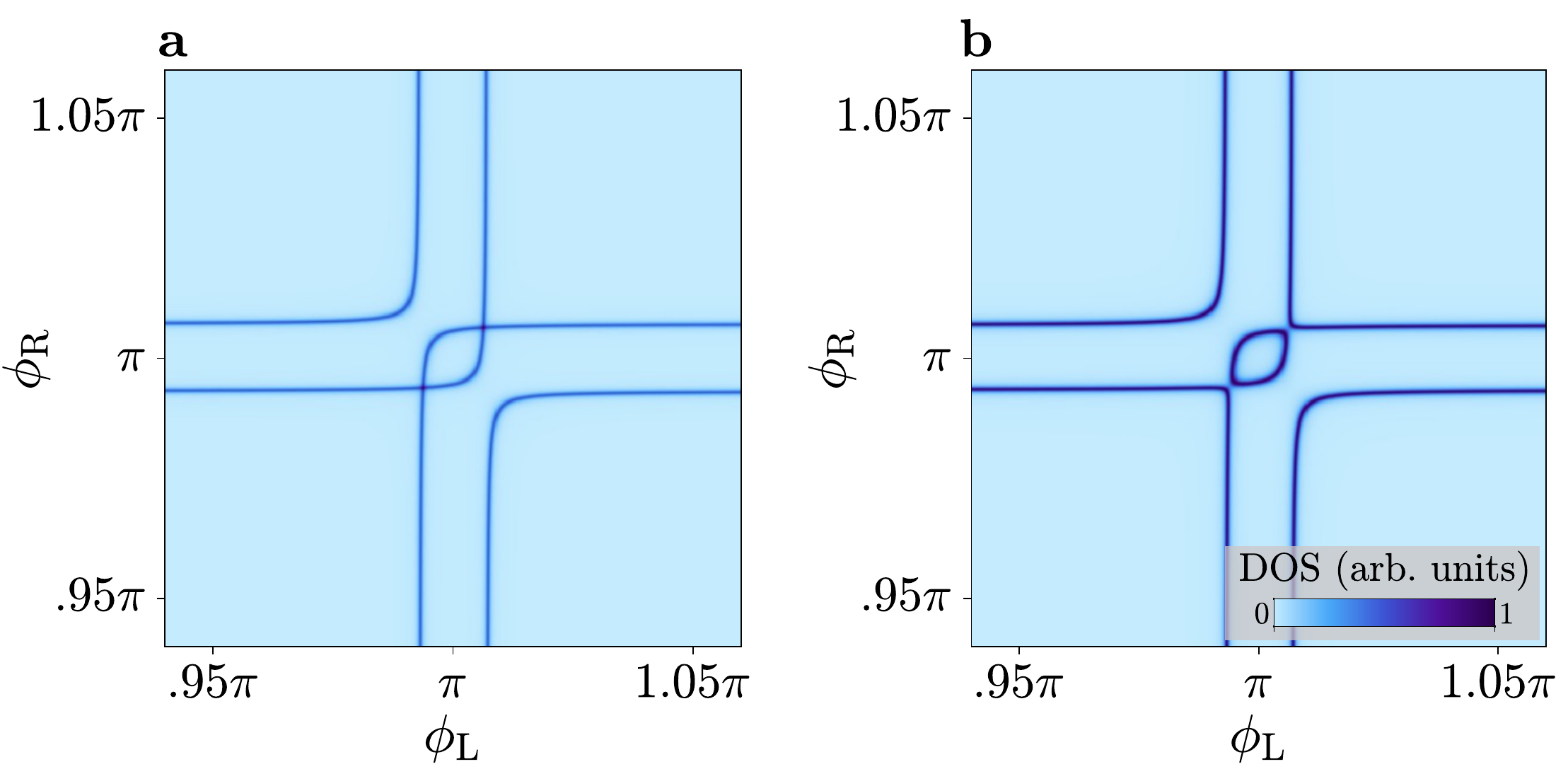}
	\caption{\textbf{Numerical simulations.} 
		\textbf{a},\textbf{b}, Density of states at fixed energy $E=-0.01 \Deltait$ as a function of the superconducting phase differences $\phiL - \phiM \equiv \phiL$ and $\phiR - \phiM \equiv \phiR$, for coupling energy $t=0.05\Deltait$ and broadening $\eta = 5 \cdot 10^{-4} \Deltait$. In \textbf{a}, $\Gammait_{\mathrm{M},1}=\Gammait_{\mathrm{M},2}=\Gammait_{\mathrm{L},1}=\Gammait_{\mathrm{R},2}=5\Deltait$, namely perfect transmissions are assumed. In \textbf{b}, $\Gammait_{\mathrm{L},1}=4.98 \Deltait$, $\Gammait_{\mathrm{M},1}=\Gammait_{\mathrm{R},2}=5 \Deltait$ and $\Gammait_{\mathrm{M},2}=4.97 \Deltait$, hence finite reflection amplitudes are considered. In both cases, the effective coupling $t/\Gammait \approx 0.01$ is similar to that introduced for the analytical case (see Fig.~\ref{figST1}).}
	\label{figST3}
\end{figure}

\setcounter{myc}{4}
\begin{figure}[p!]
	\includegraphics[width=\columnwidth]{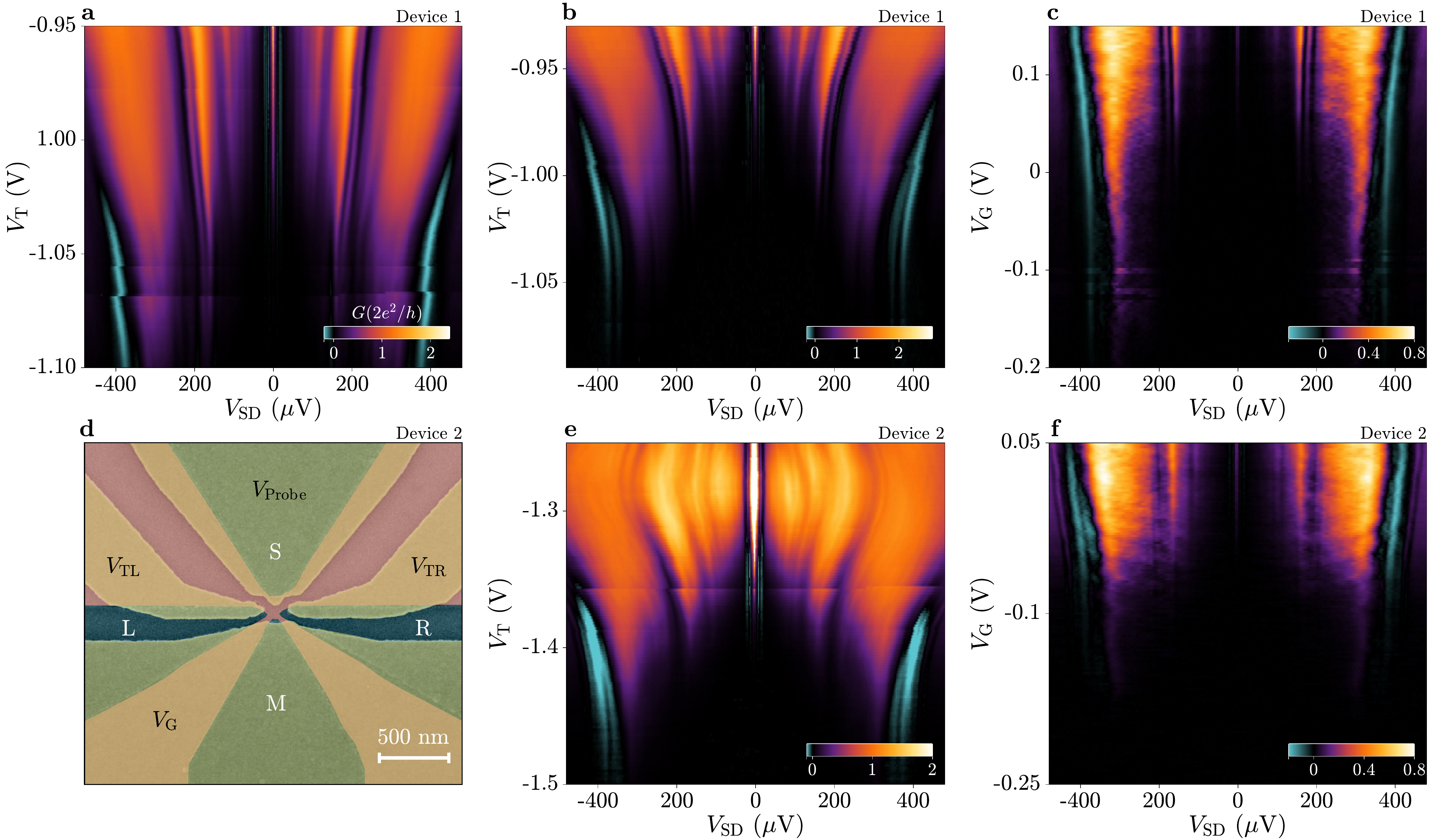}
	\caption{\textbf{Gate dependence of differential conductance.}
		\textbf{a},\textbf{b}, Differential conductance $G$ as a function of voltage bias $\Vsd$ and tunnelling gates voltage $\Vt \equiv \Vtl = \Vtr$ in Device 1, for global gate voltage $\Vg = 50~\mathrm{mV}$ in \textbf{a} and $\Vg = -150~\mathrm{mV}$ in \textbf{b}.
		\textbf{c}, $G$ as a function of $\Vsd$ and $\Vg$ in Device 1, for $\Vt = -1.07~\mathrm{V}$.
		\textbf{d}, False-coloured scanning electron micrograph of Device 2 in proximity of the three-terminal junction region (see Fig.~1a--c of the Main Text for the colour legend).
		\textbf{e}, $G$ as a function of $\Vsd$ and $\Vt$ in Device 2, for $\Vg = -150~\mathrm{mV}$ and $\Vp = 100~\mathrm{mV}$.
		\textbf{f}, $G$ as a function of $\Vsd$ and $\Vg$ in Device 2, for $\Vt = -1.5~\mathrm{V}$ and $\Vp = 100~\mathrm{mV}$.}
	\label{Sfig1}
\end{figure}

\clearpage

\setcounter{myc}{5}
\begin{figure}[p!]
	\includegraphics[width=\columnwidth]{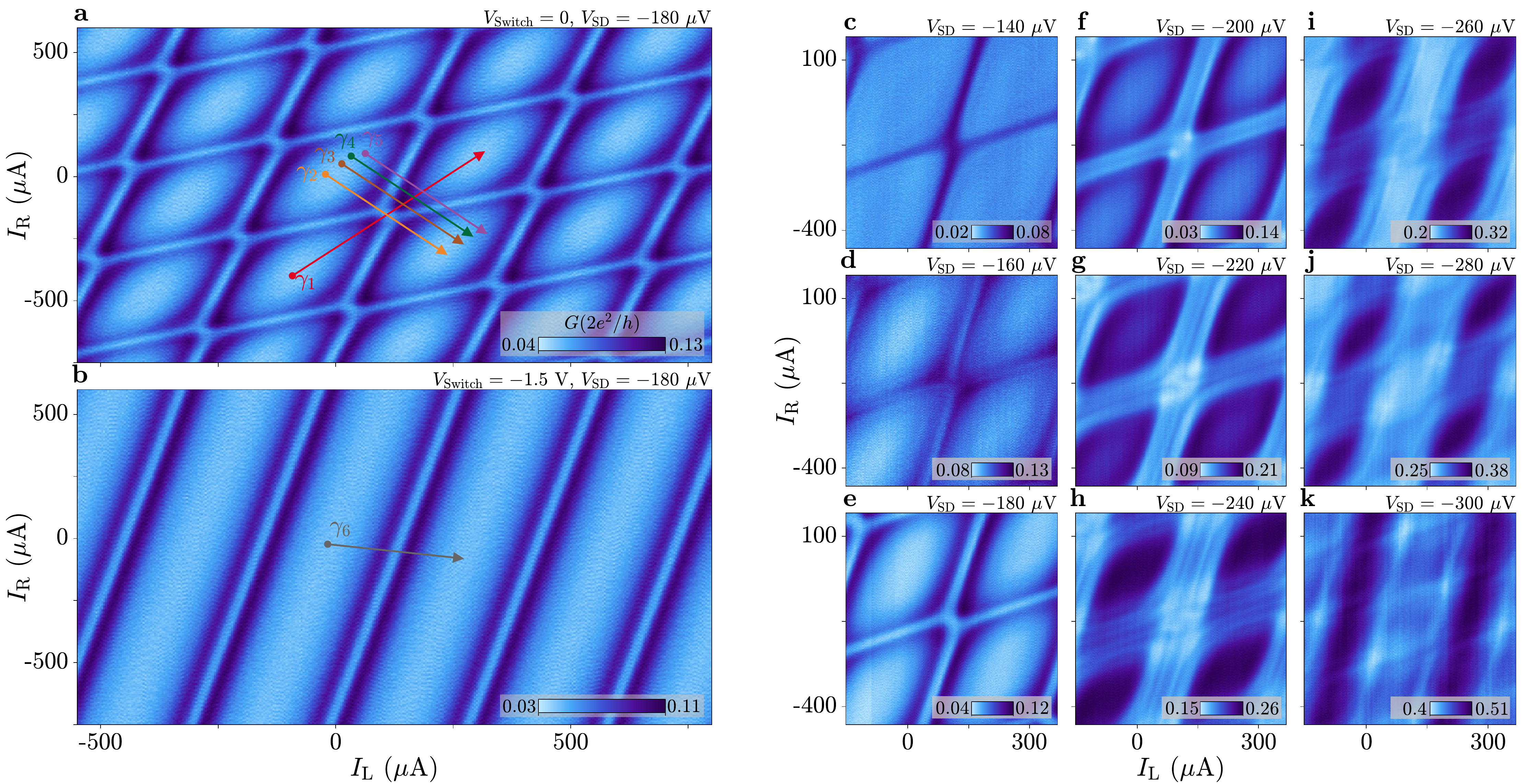}
	\caption{\textbf{Constant-bias planes as a function of the two phases in Device 2.} 
		\textbf{a}--\textbf{k}, Differential conductance $G$ as a function of the currents $\Il$ and $\Ir$ injected into the flux-bias lines at fixed values of voltage bias $\Vsd$.
		\textbf{a}, $\Vj = 0$, $\Vsd = -180~\mu \mathrm{V}$.
		\textbf{b}, $\Vj = -1.5~\mathrm{V}$, $\Vsd = -180~\mu \mathrm{V}$.
		\textbf{c}--\textbf{k}, $\Vj = 0$, $\Vsd$ varied between $-140~\mu \mathrm{V}$ (\textbf{c}) and $-300~\mu \mathrm{V}$ (\textbf{k}).}%in steps of $-20~\mu \mathrm{V}$.
	\label{figED1}
\end{figure}

\setcounter{myc}{6}
\begin{figure}[p!]
	\includegraphics[width=0.5\columnwidth]{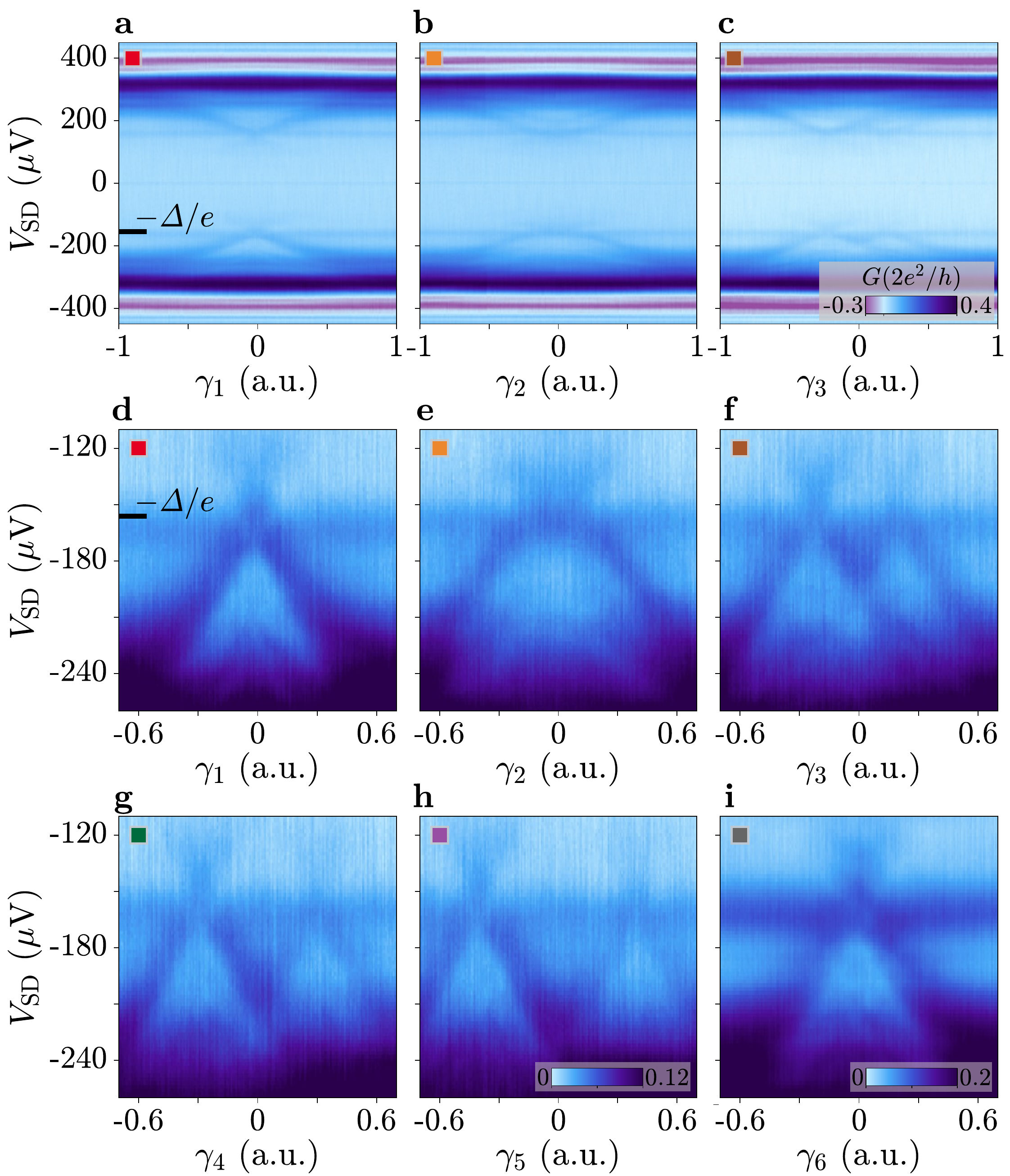}
	\caption{\textbf{Tunnelling conductance along phase space linecuts in Device 2.} 
		\textbf{a}--\textbf{c}, Differential conductance $G$ as a function of voltage bias $\Vsd$ along the linecuts $\gamma_i$ (coloured arrows in Fig.~\ref{figED1}a), for $\Vj = 0$.
		\textbf{d}--\textbf{h}, As \textbf{a}--\textbf{c}, but plotted over restricted ranges of $\Vsd$ and $\gamma_i$.
		\textbf{i}, As \textbf{d}--\textbf{h}, but along linecut $\gamma_6$ (defined in Fig.~\ref{figED2}), for $\Vj=-1.5~\mathrm{V}$. The colourbar in \textbf{h} applies to \textbf{d}--\textbf{h}.}
	\label{figED2}
\end{figure}

\setcounter{myc}{7}
\begin{figure}[p!]
	\includegraphics[width=\columnwidth]{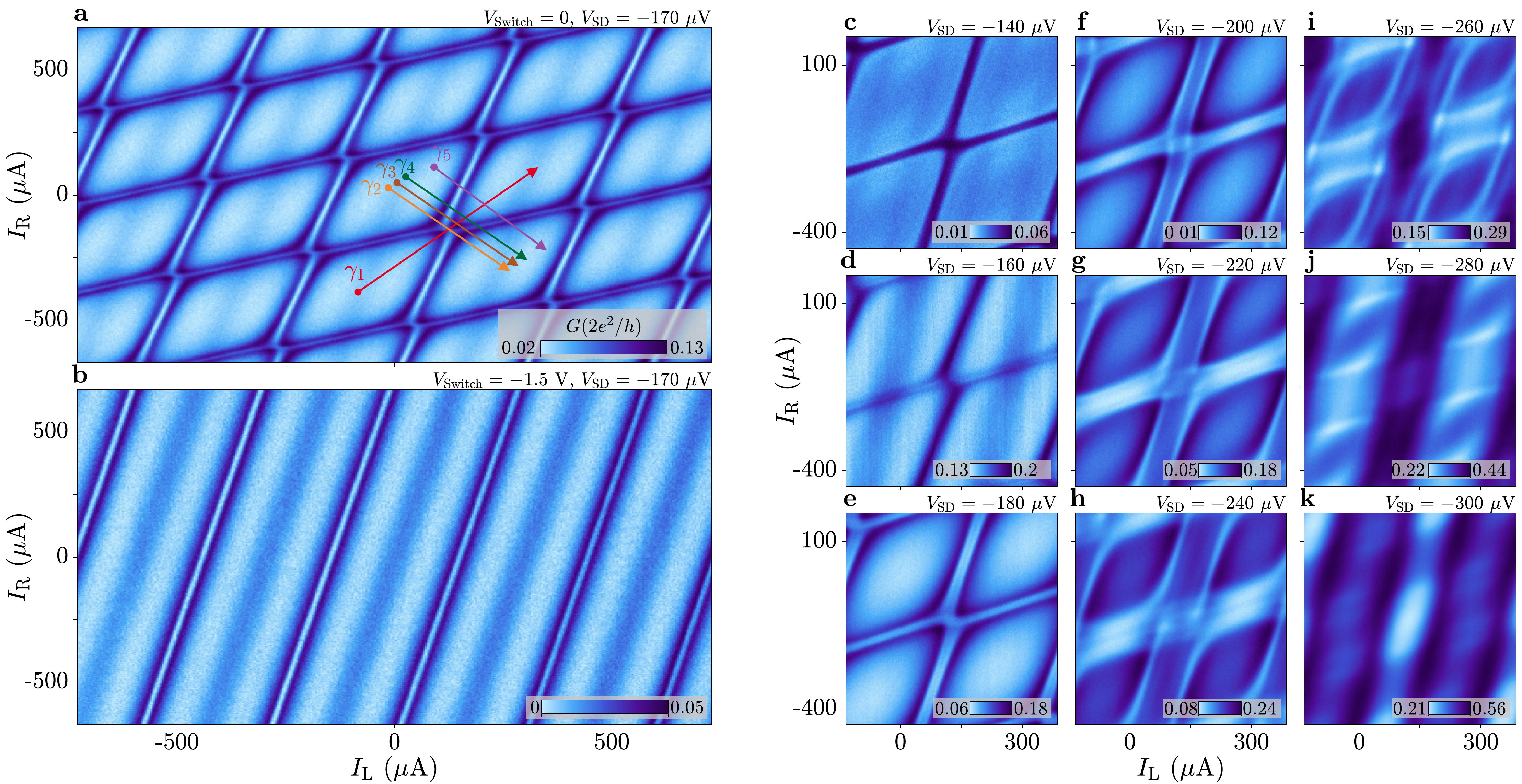}
	\caption{\textbf{Constant-bias planes as a function of the two phases at $\Vg=-66~\mathrm{mV}$.} 
		\textbf{a}--\textbf{k}, Differential conductance $G$ as a function of the currents $\Il$ and $\Ir$ injected into the flux-bias lines at fixed values of voltage bias $\Vsd$.
		\textbf{a}, $\Vj = 0$, $\Vsd = -170~\mu \mathrm{V}$.
		\textbf{b}, $\Vj = -1.5~\mathrm{V}$, $\Vsd = -170~\mu \mathrm{V}$.
		\textbf{c}--\textbf{k}, $\Vj = 0$, varying $\Vsd$.}
	\label{Sfig2}
\end{figure}

\setcounter{myc}{8}
\begin{figure}[p!]
	\includegraphics[width=0.5\columnwidth]{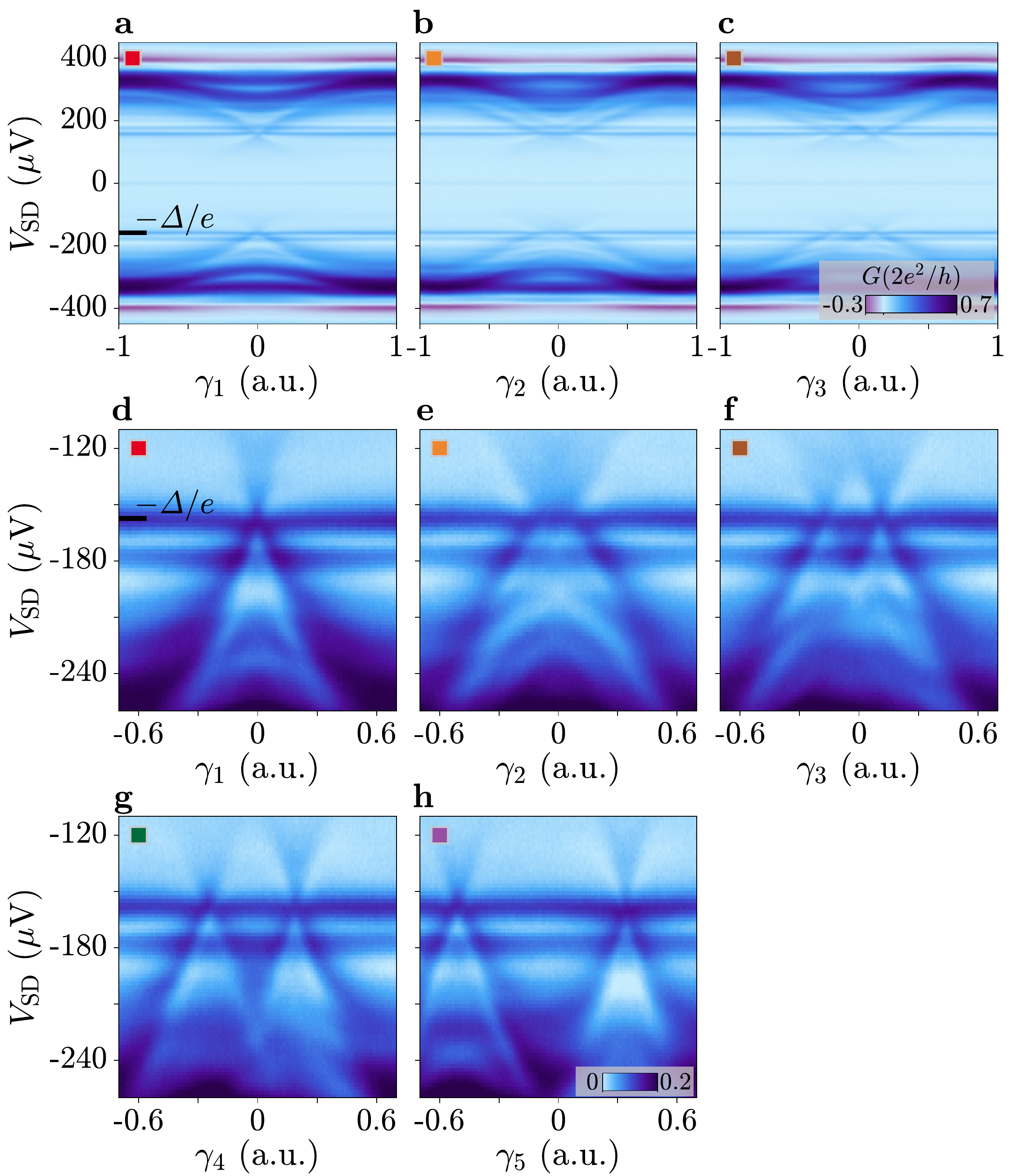}
	\caption{\textbf{Tunnelling conductance along phase space linecuts at $\Vg=-66~\mathrm{mV}$.} 
		\textbf{a}--\textbf{c}, Differential conductance $G$ as a function of voltage bias $\Vsd$ along the linecuts $\gamma_i$ (coloured arrows in Fig.~\ref{Sfig2}a), for $\Vj = 0$.
		\textbf{d}--\textbf{h}, As \textbf{a}--\textbf{c}, but plotted over restricted ranges of $\Vsd$ and $\gamma_i$.}
	\label{Sfig3}
\end{figure}

\setcounter{myc}{9}
\begin{figure}[p!]
	\includegraphics[width=\columnwidth]{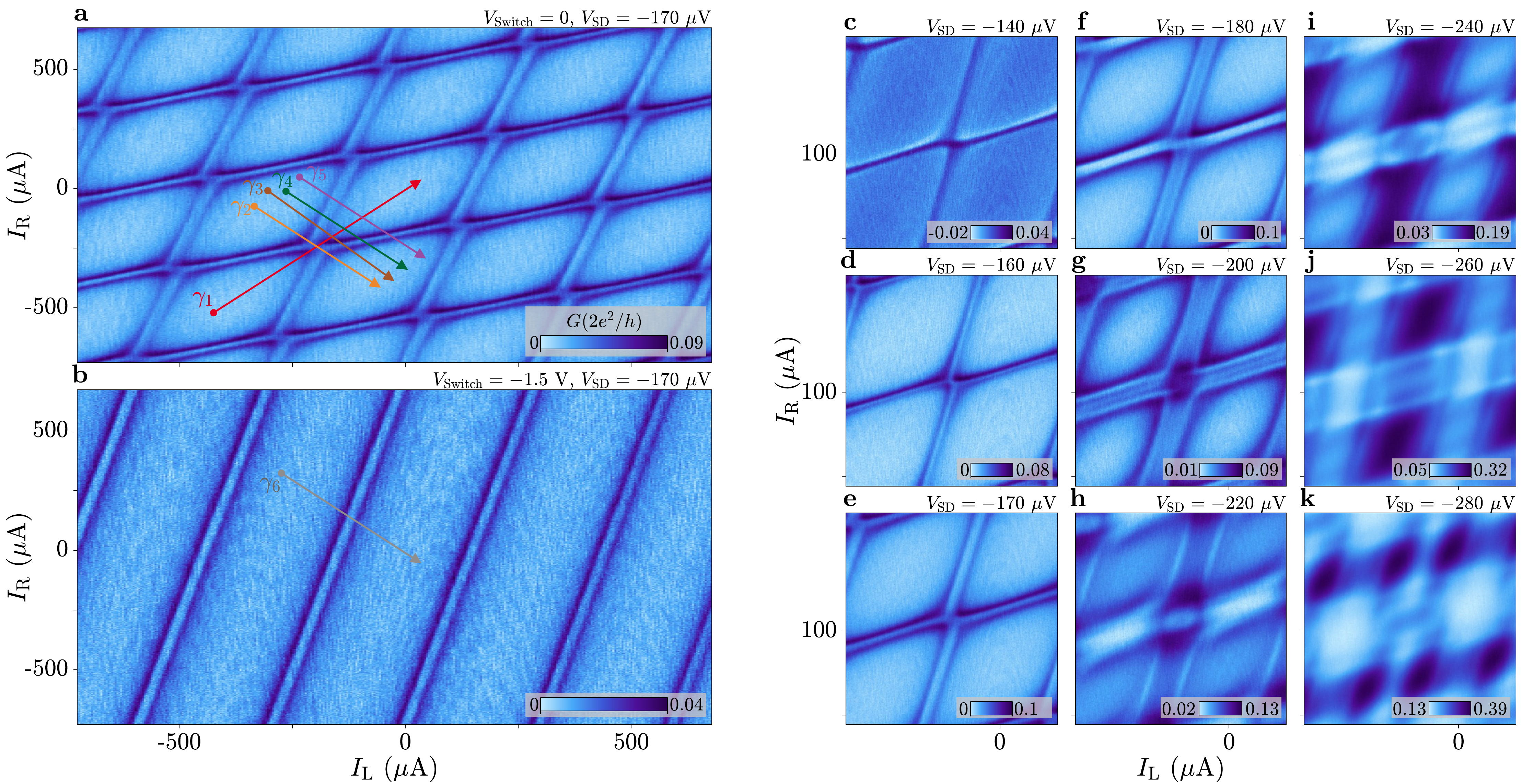}
	\caption{\textbf{Constant-bias planes as a function of the two phases at $\Vg=-150~\mathrm{mV}$.} 
		\textbf{a}--\textbf{k}, Differential conductance $G$ as a function of the currents $\Il$ and $\Ir$ injected into the flux-bias lines at fixed values of voltage bias $\Vsd$.
		\textbf{a}, $\Vj = 0$, $\Vsd = -170~\mu \mathrm{V}$.
		\textbf{b}, $\Vj = -1.5~\mathrm{V}$, $\Vsd = -170~\mu \mathrm{V}$.
		\textbf{c}--\textbf{k}, $\Vj = 0$, varying $\Vsd$.}
	\label{Sfig4}
\end{figure}

\setcounter{myc}{10}
\begin{figure}[p!]
	\includegraphics[width=0.5\columnwidth]{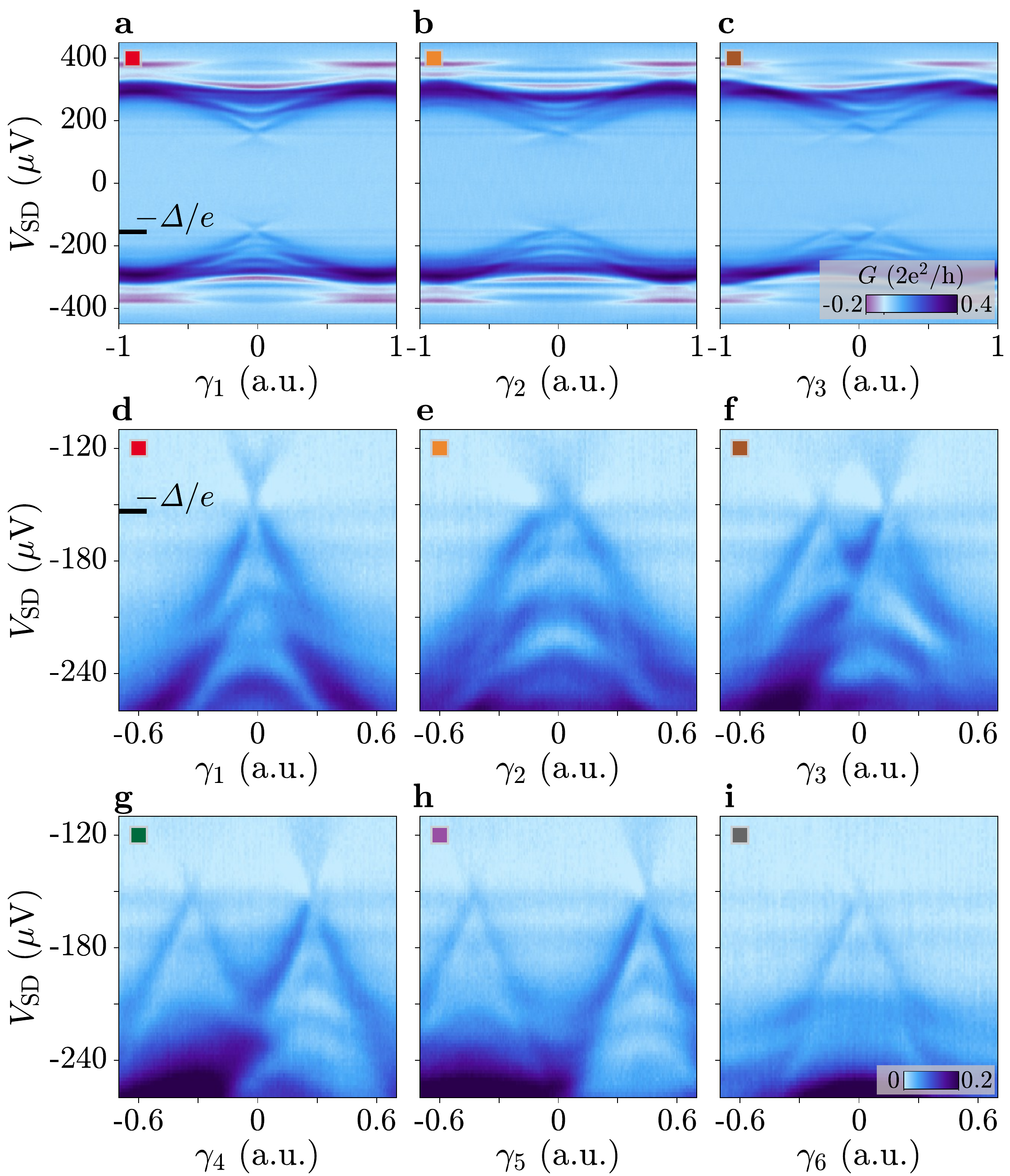}
	\caption{\textbf{Tunnelling conductance along phase space linecuts at $\Vg=-150~\mathrm{mV}$.} 
		\textbf{a}--\textbf{c}, Differential conductance $G$ as a function of voltage bias $\Vsd$ along the linecuts $\gamma_i$ (coloured arrows in Fig.~\ref{Sfig4}a), for $\Vj = 0$.
		\textbf{d}--\textbf{h}, As \textbf{a}--\textbf{c}, but plotted over restricted ranges of $\Vsd$ and $\gamma_i$.
		\textbf{i}, As \textbf{d}--\textbf{h}, but along linecut $\gamma_6$ (defined in Fig.~\ref{Sfig4}b), for $\Vj=-1.5~\mathrm{V}$.}
	\label{Sfig5}
\end{figure}

\setcounter{myc}{11}
\begin{figure}[p!]
	\includegraphics[width=\columnwidth]{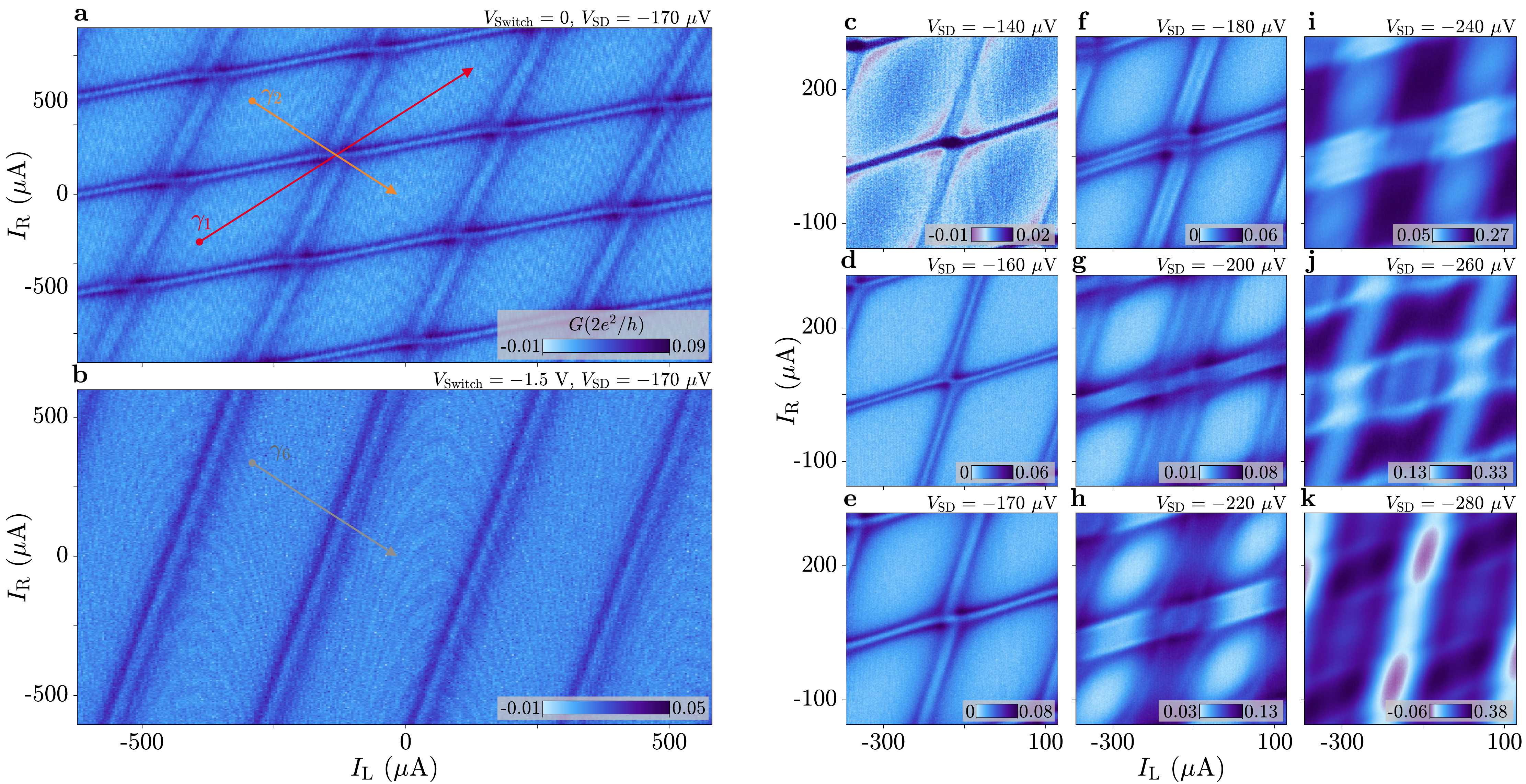}
	\caption{\textbf{Constant-bias planes as a function of the two phases at $\Vg=-250~\mathrm{mV}$.} 
		\textbf{a}--\textbf{k}, Differential conductance $G$ as a function of the currents $\Il$ and $\Ir$ injected into the flux-bias lines at fixed values of voltage bias $\Vsd$.
		\textbf{a}, $\Vj = 0$, $\Vsd = -170~\mu \mathrm{V}$.
		\textbf{b}, $\Vj = -1.5~\mathrm{V}$, $\Vsd = -170~\mu \mathrm{V}$.
		\textbf{c}--\textbf{k}, $\Vj = 0$, varying $\Vsd$.}
	\label{Sfig6}
\end{figure}

\setcounter{myc}{12}
\begin{figure}[p!]
	\includegraphics[width=0.5\columnwidth]{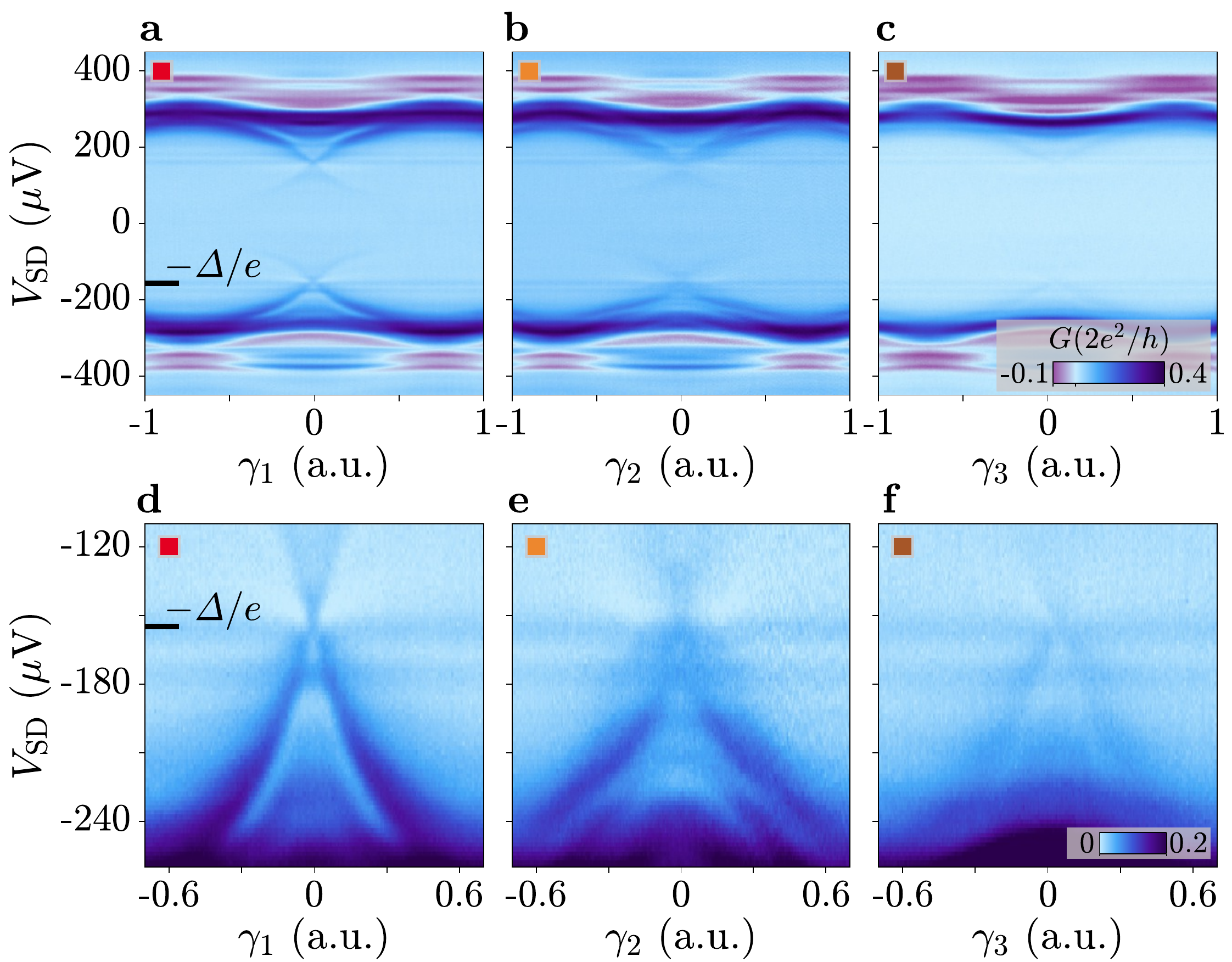}
	\caption{\textbf{Tunnelling conductance along phase space linecuts at $\Vg=-250~\mathrm{mV}$.} 
		\textbf{a}--\textbf{c}, Differential conductance $G$ as a function of voltage bias $\Vsd$ along the linecuts $\gamma_i$ (coloured arrows in Fig.~\ref{Sfig6}a,b). $\Vj = 0$ in \textbf{a}, \textbf{b} and $\Vj = -1.5~\mathrm{V}$ in \textbf{c}.
		\textbf{d}--\textbf{f}, As \textbf{a}--\textbf{c}, but plotted over restricted ranges of $\Vsd$ and $\gamma_i$.}
	\label{Sfig7}
\end{figure}

\setcounter{myc}{13}
\begin{figure}[p!]
	\includegraphics[width=0.5\columnwidth]{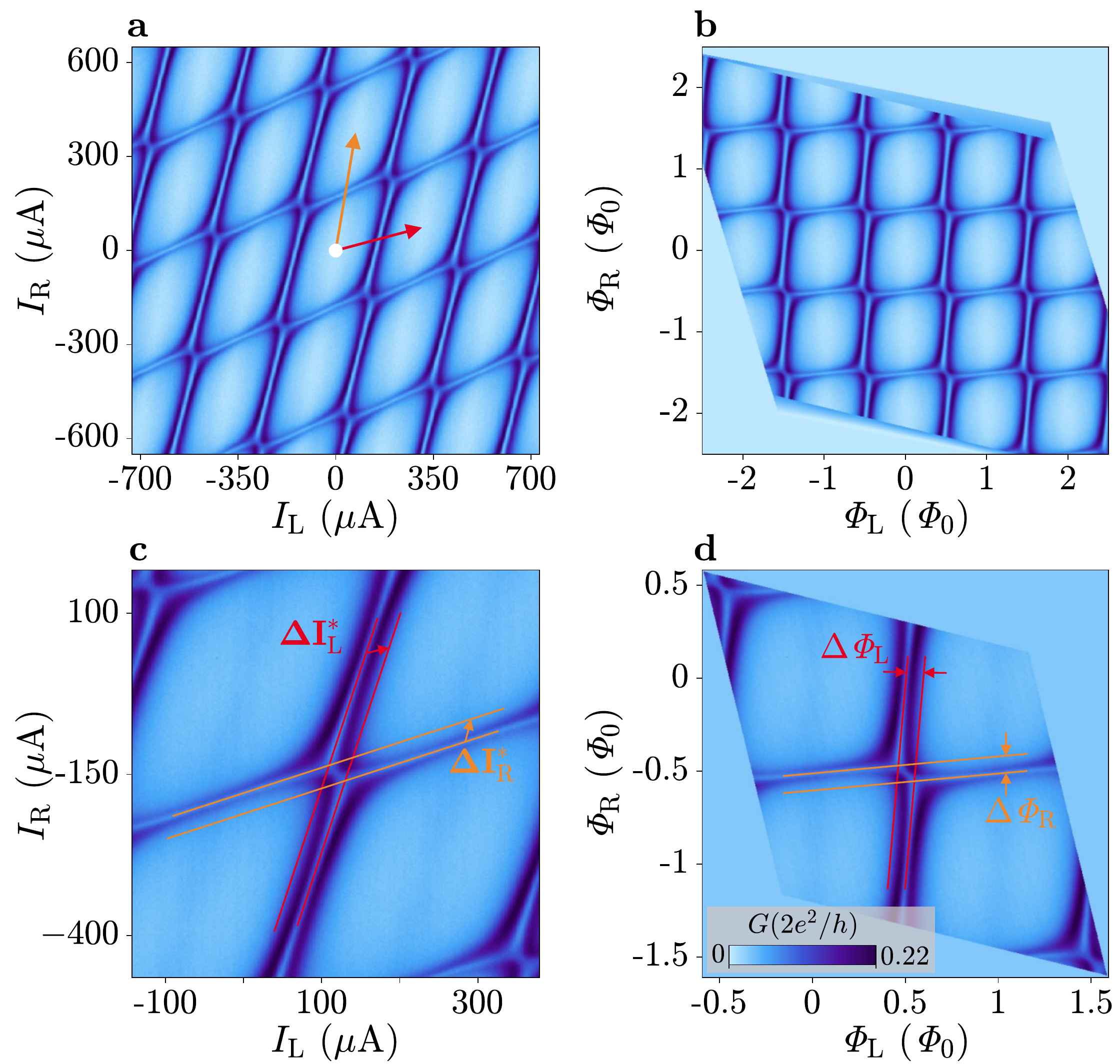}
	\caption{\textbf{Current-to-flux remapping and phase shift analysis.}
		\textbf{a}, Differential conductance $G$ as a function of the currents $\Il$ and $\Ir$ injected into the flux-bias lines at fixed voltage bias $\Vsd = -170 ~\mu \mathrm{V}$ (repetition of Fig.~1d from the Main Text). The red and orange arrows indicate the periodicity axes, namely the directions of the external magnetic flux axes $\PhiL$ and $\PhiR$. The white dot and the end points of the red and orange arrows correspond to the $(\PhiL, \PhiR)$ points $(0,0)$, $(\mathit{\Phi}_0, 0)$ and $(0, \mathit{\Phi}_0)$, respectively.
		\textbf{b}, As \textbf{a}, but remapping the $(\Il, \Ir)$ axes to the $(\PhiL, \PhiR)$ axes by using Eqs.~\ref{eqM} and \ref{eqM2} (see text for details).
		\textbf{c}, $G$ as a function of $\Il$ and $\Ir$ at $\Vsd = -165 ~\mu \mathrm{V}$ (repetition of Fig.~4e from the Main Text). The red (orange) lines follow the middle dip between the two resonances of the $\mathit{\Phi}_\mathrm{L(R)}$-dependent Andreev bound state, on both sides of the intersection point. The $\mathit{\Phi}_\mathrm{L(R)}$ shift is indicated by the vector $\mathbf{\Delta I}^*_\mathrm{L(R)}$, whose direction is parallel to the $\mathit{\Phi}_\mathrm{L(R)}$ axis on the $(\Il, \Ir)$ plane (see \textbf{a}).
		\textbf{d}, As \textbf{c}, but remapping the $(\Il, \Ir)$ axes to the $(\PhiL, \PhiR)$ axes. The phase shifts $\Delta \PhiL$ and $\Delta \PhiR$, by construction parallel to the axes, are indicated by the arrows.}
	\label{Sfig8}
\end{figure}

\end{document}